\def\fulltitle{Learning sequence timing and control of replay speed in networks of spiking neurons}
\def\shorttitle{Sequence timing and replay speed}
\def\fullauthors{Melissa Lober, Younes Bouhadjar, Markus Diesmann, Tom Tetzlaff}
\def\shortauthors{Lober et al.}
\def\figscale{0.85}

\documentclass[10pt,a4paper,twoside,american]{article}

\usepackage[
breaklinks=true,
colorlinks=true,
linkcolor=blue,citecolor=blue,urlcolor=blue,filecolor=blue,
pdffitwindow,
bookmarks=true,
bookmarksopen=true,
bookmarksnumbered=true,
pdftitle={\fulltitle},
pdfauthor={\fullauthors},
pdfsubject={},
pdfkeywords={}
]{hyperref}  

\usepackage{amsmath}
\usepackage{amssymb}
\usepackage[toc,page,titletoc]{appendix}
\usepackage{authblk}
\usepackage{bm}
\usepackage{calc}
\usepackage[font=small,labelfont=bf]{caption}
\usepackage{float}
\usepackage{geometry}
\usepackage{graphicx}
\usepackage{lineno}
\usepackage{mathtools}
\usepackage{nameref}
\usepackage[square,sort,comma,numbers]{natbib}
\usepackage{refstyle}
\newref{sec}{refcmd=section~\hyperref[#1]{\nameref{#1}}}
\usepackage[table]{xcolor}
\definecolor{shadecolor}{gray}{0.9}

\usepackage{cleveref}
\crefname{supp}{supplement}{Supplements}
\crefname{app}{appendix}{Appendices}
\crefformat{equation}{Eq.\,(#2#1#3)}
\crefformat{figure}{Fig.\,#2#1#3}
\crefformat{appendix}{App.\,#2#1#3}
\crefformat{table}{Tab.\,#2#1#3}
\crefformat{video}{Video\,#2#1#3}
\crefformat{algorithm}{Algorithm\,#2#1#3}
\crefformat{section}{Sec.\,#2#1#3}

\usepackage{changes}        

\definecolor{TT}{RGB}{0,200,0}
\definecolor{ML}{RGB}{200,0,200}
\definecolor{YB}{RGB}{0,150,255}
\definecolor{MD}{RGB}{200,150,150}
\definecolor{REV}{RGB}{220,0,0}

\definechangesauthor[color=TT]{TT}
\definechangesauthor[color=ML]{ML}
\definechangesauthor[color=YB]{YB}
\definechangesauthor[color=MD]{MD}
\definechangesauthor[color=REV]{REV}

\setcommentmarkup{\color{authorcolor}{}[#1]}  

\usepackage{fancyhdr}
\renewcommand{\secref}[1]{``\nameref{#1}''} 

\makeatletter
\renewcommand{\@biblabel}[1]{\quad#1.}
\makeatother

\newcommand{\CM}{C_\textnormal{m}}    
\newcommand{\derived}[1]{{\color{gray}{}#1}}

\newcommand{\dtsim}{\Delta t}

\newcommand{\EE}{{\exc\exc}}
\newcommand{\EI}{{\exc\inh}}

\newcommand{\exc}{\textnormal{E}}     
\newcommand{\ext}{\textnormal{X}}     

\newcommand{\IE}{{\inh\exc}}

\newcommand{\ET}{{\exc\textnormal{T}}}
\newcommand{\EX}{{\exc\ext}}
\newcommand{\Epop}{\mathcal{E}} 
\newcommand{\inh}{\textnormal{I}}     
\newcommand{\Ipop}{\mathcal{I}} 

\newcommand{\J}{J}                          

\newcommand{\JEI}{\J_{\exc\inh}}

\newcommand{\JIE}{\J_{\inh\exc}}

\newcommand{\JET}{{\J_{\exc\textnormal{T}}}}
\newcommand{\JEX}{{\J_{\exc\ext}}}

\newcommand{\KEE}{{K_{\exc\exc}}}

\newcommand{\mat}[1]{\bm{#1}}
\newcommand{\ms}{\,\textnormal{ms}}

\newcommand{\mV}{\,\textnormal{mV}}

\newcommand{\NE}{{N_{\exc}}}

\newcommand{\nE}{{n_\exc}}
\newcommand{\nI}{{n_\inh}}
\newcommand{\pA}{\,\textnormal{pA}}

\newcommand{\pF}{\,\textnormal{pF}}

\newcommand{\s}{\,\textnormal{s}}

\newcommand{\seq}[1]{\ensuremath{\{\text{#1}\}}}

\newcommand{\Tpop}{\mathcal{T}}

\renewcommand{\vec}[1]{\bm{#1}}
\newcommand{\Vreset}{V_\textnormal{r}}

\newcommand{\Xpop}{\mathcal{X}}

\title{\fulltitle} 

\author[1,2,*]{Melissa Lober}
\author[3,4]{Younes Bouhadjar}
\author[1,5,6]{Markus Diesmann}
\author[1]{Tom Tetzlaff}

\affil[1]{\footnotesize%
  Institute for Advanced Simulation (IAS-6), J\"ulich Research Centre, J\"ulich, Germany}
\affil[2]{\footnotesize%
  RWTH Aachen University, Aachen, Germany}
\affil[3]{\footnotesize%
  Peter Gr\"unberg Institute, Neuromorphic Software Ecosystems (PGI-15), J\"ulich Research Centre, Germany}
\affil[4]{\footnotesize%
  Fraunhofer IIS, Erlangen, Germany}
\affil[5]{\footnotesize%
  Department of Physics, Faculty 1, \& Department of Psychiatry, Psychotherapy, and Psychosomatics, Medical School, RWTH Aachen University, Aachen, Germany}
\affil[6]{\footnotesize%
JARA BRAIN Institute Structure-Function Relationships (INM-10), J\"ulich Research Centre, J\"ulich, Germany}
  
\affil[*]{\footnotesize\url{m.lober@fz-juelich.de}}

\date{\footnotesize\today}

\begin{document}

\maketitle
\pdfbookmark[1]{Title}{title}
\pagestyle{fancy}


\begin{abstract}
Processing sequential inputs is a fundamental brain function, underlying tasks such as sensory perception, language, and motor control. 
A challenge in sequence processing is to represent not only the order of events, but also their precise timing.
While existing computational models can learn sequential structure, many lack biologically plausible mechanisms to encode element-specific timing and to flexibly control the speed of sequence replay.
The spiking Temporal Memory (sTM) model, a biologically inspired network model, provides a framework for key aspects of sequence processing.
In the sTM model, each sequence element is represented by a small set of neurons firing synchronously, where the set of active neurons encodes the element's identity in its sequential context.
In its original version, however, the sTM model learns the order but not the timing of sequence elements. 
Further, it remains an open question in neuroscience how the speed of sequence replay can be flexibly modulated.
We propose a mechanism where the duration of sequence elements is represented by a sequential activation of element specific neuronal populations, enabling the model to encode sequences across a wide range of timescales.
This provides a biologically plausible basis for learning and replaying complex temporal patterns.
Additionally, we show that oscillatory background inputs can serve as a clock signal and provide a robust and flexible mechanism for controlling the speed of sequence replay.
Our findings suggest that elapsed time is encoded by unique and sparse spatiotemporal patterns of neural activity, and that the speed of sequence replay during wakefulness and sleep is correlated to the characteristics of global oscillatory activity observed in EEG or LFP recordings.
\end{abstract}
\section{Author summary}
Sequences are everywhere in our daily lives. We encounter them when we speak, listen to music, move our bodies, or interpret the visual world around us. We can replay learned sequences at different speeds while preserving their rhythm. A melody, for example, can be played slowly during practice or quickly during performance, and still remain recognizable, showing that the brain must represent both the order of events and their timing.
In this study, we use a biologically inspired model to explore how the brain might encode the timing of events and flexibly replay learned sequences at different speeds. Our results suggest a mechanism in which sequential patterns of neural activity and ongoing brain rhythms work together to control the pace of sequence replay. These findings provide new insight into how the brain may organize the timing of thoughts and actions, and lead to testable predictions about how neural activity and brain oscillations interact during sequence replay.
\section{Introduction}
Learning and processing sequences of events is a fundamental computational principle of the brain \cite{Lashley51_112,Clegg98_275,Hawkins07_on_intelligence,Dehaene15_2}. During perception, we continuously interpret temporally structured input such as speech, music, or complex visual scenes. During action, we generate precisely timed motor sequences, from locomotion to speech production. Across these domains, sequence processing involves several tightly coupled operations: forming context-dependent predictions of upcoming events, detecting deviations from expected input, and replaying previously learned sequences in response to partial cues. These capabilities are central to cognition and are thought to be fundamental functions of the neocortex.
An essential aspect of sequence processing is time. Natural sequences are not only defined by the order of their elements, but also by their durations and temporal structure \cite{Levitin96,Boltz10}. Musical melodies, spoken language, and motor behaviors all rely on precisely controlled timing. Moreover, behavioral experience indicates that learned sequences can later be replayed at different speeds. For example, motor sequences can be executed slowly during practice and rapidly during performance, and episodic memories can be mentally replayed in compressed or expanded form. At the same time, this temporal flexibility is not unlimited, suggesting that neural mechanisms impose constraints on both learning and replay of temporal structure. Despite extensive experimental and theoretical work on sequence learning, the mechanisms that allow neural circuits to encode element durations and flexibly modulate replay speed remain unclear.
While modern sequence models such as Transformers or deep recurrent networks can learn to predict and reproduce sequential inputs successfully, they rely on dense representations, global learning rules, and positional encodings with no identified counterpart in neural circuitry. Understanding how the brain solves sequence processing therefore requires a different approach, grounded in sparse population codes, local synaptic plasticity, and intrinsic temporal dynamics that characterize cortical computation.
\par
A broad range of biologically motivated models has been proposed to account for sequence learning and replay. Recurrent spiking network models demonstrate how synaptic plasticity gives rise to sequential assembly activity, context-dependent responses, and the learning of structured spatiotemporal patterns \cite{Klampfl13_11515,Klos18_e1006187,Maes20_e1007606,Cone21_e63751,Asabuki22,Calderon22,Kriener24_elise}. While these models capture important aspects of sequential processing, they typically focus on either learning or replay, and do not simultaneously address context-dependent prediction, anomaly detection, and autonomous replay within a single unified framework. The Temporal Memory algorithm of the Hierarchical Temporal Memory (HTM) model \cite{Hawkins11_whitepaper,Hawkins16_23} provides such a unified framework. The spiking continuous-time formulation of this model (spiking Temporal Memory (sTM) model \cite{Bouhadjar22_e1010233}) links sequence learning to biologically interpretable mechanisms, such as nonlinear dendritic integration and dendritic action potentials (dAPs) \cite{Antic10_2991}, sparse population coding \cite{Barth12_345}, lateral inhibition \cite{Mullner15_576}, and structural spike-timing-dependent plasticity \cite{Liao95_400,Luescher2000_545,Deger12_e1002689}. The resulting architecture forms sequence-specific subnetworks that support context-dependent prediction, mismatch signaling, and autonomous replay. Extensions of the model further demonstrate that it can generate probabilistic replay shaped by coherent network activity \cite{Bouhadjar23_e1010989} and learn sequences of dozens of elements \cite{Bouhadjar25_202}.
\par
While these studies established the sTM model as a biologically motivated model for learning and replaying the order of sequence elements, it remains unclear whether it can learn temporal features such as the duration of sequence elements and inter-element intervals, and how the speed of replay can be modulated. The original formulation predicts bounds on replay speed that arise from synaptic and neuronal time constants \cite{Bouhadjar22_e1010233}, yet it cannot predict sequences whose elements are separated by intervals exceeding these intrinsic timescales, and the replay speed depends only on intrinsic time constants rather than on the temporal structure present during learning. The replay speed in the original sTM model can be influenced by parameters such as the neuronal firing threshold, which determine how readily dendritic events trigger somatic spikes. However, replay was demonstrated only for a specific parameter regime in which sequences were replayed at a high intrinsic speed, close to the upper limit set by neuronal and synaptic time constants. Thus, a mechanism that enables robust and flexible modulation of the replay speed over a broader range, without reconfiguring intrinsic cellular parameters, remains to be established. Experimental observations in cortex and hippocampus indicate that replay can occur at different compression factors depending on behavioral state \cite{Ji07,Euston07,Liu19,Buch21}, and psychophysical studies show that humans flexibly rescale temporal sequences while preserving relative timing \cite{Hardy18}, suggesting the presence of such modulatory control mechanisms.
\par
The present study addresses both issues within a unified spiking TM framework. We propose a mechanism that enables the learning and replay of sequences with complex rhythmic structure, including long and heterogeneous temporal intervals, while preserving sparse, context-dependent coding. In addition, we propose that oscillatory background input can serve as a flexible clock signal for replay. Oscillations are ubiquitous in cortical and hippocampal circuits and have long been hypothesized to provide temporal reference frames or clock-like signals for neural computation \cite{Fries05_474,Buzsaki06}. This mechanism suggests a functional role of network oscillations in controlling sequence dynamics.
\par
The study is organized as follows. In section \secref{sec:sequence_timing}, we show how the sTM model successfully learns sequences with complex temporal structure. In section \secref{sec:replay_speed}, we demonstrate how replay speed can be modulated through oscillatory background input and characterize the limits and robustness of this modulation. Section \secref{sec:testable_predictions} derives experimentally testable predictions that follow from the proposed framework.
The \secref{sec:discussion} relates the proposed mechanisms to experimental findings in cortex and hippocampus, compares the temporal properties of the sTM model to other biologically motivated sequence learning frameworks, summarizes limitations, and outlines directions for future extensions.
The \secref{sec:methods} provide a detailed description of the task, the network architecture, model parameters, and performance measures. 
\section{Results}
\label{sec:results}
\subsection{The spiking Temporal Memory (sTM)  model}
This section provides an overview of the key components of the sTM model \cite{Bouhadjar22_e1010233} and its adaptations \cite{Bouhadjar25_202} used in this study.
A detailed mathematical description of the model and the full set of parameters are given in \secref{sec:methods}.
\par
\begin{figure}[ht] 
  \centering
  \includegraphics[scale=\figscale]{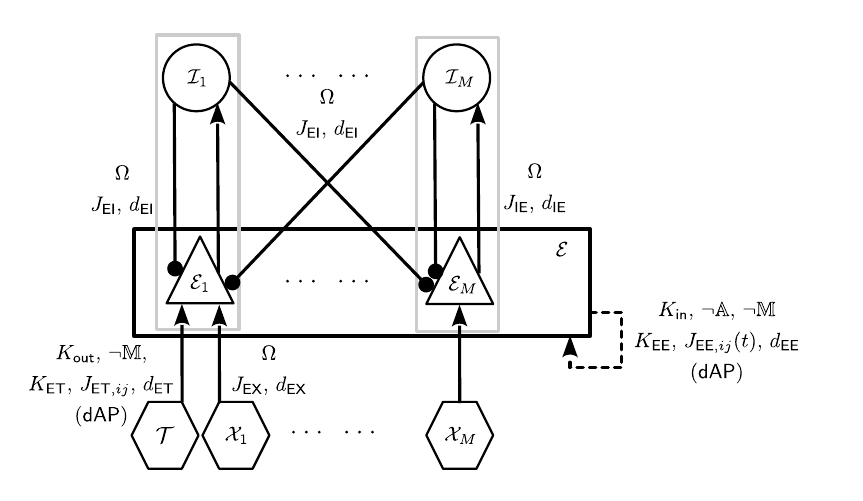}
  \caption{
    \textbf{Sketch of the sTM network architecture.}
    Graphical description of network according to \cite{Senk22_e1010086}.
    \textit{Populations:}
    $\mathcal{E}_k$: excitatory populations.
    $\mathcal{I}_k$: inhibitory populations.
    $\mathcal{X}_k$: bottom-up inputs ($k=1,\ldots,M$).    
    $\mathcal{T}$: top-down input ($\mathcal{E}_1$ hosts the assembly representing the first element of the input sequence). 
    \textit{Connectivity rules:}    
    Triangular arrow heads: excitatory connections.
    Circular arrow heads: inhibitory connections.
    Solid lines: static connections.
    Dashed lines: plastic connections.
    $K_\text{in}$: random connectivity with fixed indegree.
    $K_\text{out}$: random connectivity with fixed outdegree.
    $\Omega$: all-to-all connectivity.
    $\neg\mathbb{M}$: multiple connections between two neurons not permitted.
    $\neg\mathbb{A}$: self-connections not permitted.
    Recurrent connections between excitatory neurons (dashed) target distal dendritic branches and may trigger dendritic action potentials (dAPs) signaling predictions.
    \textit{Connectivity parameters:}
    synaptic weights $J_{xy}$,
    spike transmission delays $d_{xy}$ ($x,y\in\{\text{E},\text{I},\text{X},\text{T}\}$).
    See \secref{sec:methods} for detailed mathematical description.
  }
  \label{fig:network_sketch}
\end{figure}

The sTM is a randomly and sparsely connected recurrent neural network (\cref{fig:network_sketch}) organized into $M$ subpopulations (minicolumns, \cite{Mountcastle97_701,Buxhoeveden02_935}), each consisting of $n_\text{E}$ excitatory ($\mathcal{E}_k$; $k=1,\ldots,M$) and $n_\text{I}$ inhibitory neurons ($\mathcal{I}_k$; here:  $M=6$, $n_\text{E}=200$, $n_\text{I}=1$).
Excitatory neurons within each minicolumn project to their associated inhibitory neuron, which in turn inhibits all excitatory neurons in the network.
On the one hand, this architecture enforces a competition between minicolumns \cite{Lundqvist06_253,Johansson07_1871}. On the other hand, it implements a local winner-take-all (WTA) mechanism that promotes sparse activity \cite{Maass00_2519}.
During sequence presentation, excitatory neurons within each minicolumn receive stimulus-specific, suprathreshold bottom-up inputs ($\mathcal{X}_k$).
Lateral excitatory connections between excitatory neurons target distal dendritic branches.
When the synaptic input to a dendritic branch exceeds a threshold, a dendritic action potential (dAP; ``NMDA spike'') is generated, leading to a long-lasting ($100\,\text{ms}$) subthreshold plateau depolarization of the soma \cite{Antic10_2991}.
These plateau potentials signal predictions: after successful learning, they precede the arrival of an expected bottom-up input. When the expected input arrives, predictive neurons fire earlier than others, confirming the prediction. If a sufficient number of $\rho$ neurons within a minicolumn are in a predictive state and hence fire earlier, the local inhibitory feedback kicks in and prevents all other (late) neurons in the minicolumn from firing (WTA). This selection of a sparse subset of active neurons leads to a context-specific representation of the current sequence element.
After learning, the same input activates sequence (context) specific neuronal groups, here referred to as assemblies \cite{Palm82}.
These assemblies are defined by their consistent co-activation rather than by stronger recurrent connectivity within the group \cite{Wagner-Carena25}.
This produces a temporally evolving pattern of context-specific sparse activity (``bar code'') that represents the progression of the sequence (see Fig.~\ref{fig:slow-sequences}A).
In the absence of any additional information, the first element in a sequence can not be anticipated. A bottom-up input representing the first element in the sequence leads to a non-sparse response of all neurons in the corresponding minicolumn, and therefore to an ambiguous activation of all possible assemblies present in this minicolumn.
Here, we avoid this ambiguity by assuming that the start of a given sequence is predicted by an additional top-down input ($\mathcal{T}$), which activates the dendrites of a sparse subset of neurons in the corresponding minicolumn prior to the arrival of the bottom-up stimulus representing the first sequence element.
Upon arrival of this stimulus, only this sparse subset of neurons fires.
Synapses between excitatory neurons are plastic and evolve according to a structural spike-timing-dependent plasticity rule combined with a continuous synaptic weight decay (``forgetting'').
This plasticity mechanism was first introduced in \cite{Bouhadjar25_202}, and differs from the original formulation \cite{Bouhadjar22_e1010233}.
\subsection{Learning sequence timing}
\label{sec:sequence_timing}
A key temporal constraint of the sTM model arises from the finite duration of the plateau potentials, typically lasting few hundred milliseconds \cite{Schiller2000_285, Milojkovic05_3940, Larkum09_325, Antic10_2991}.
As a result, the maximum interval between two consecutive elements in a sequence that can be bridged by dendritic activity is limited.
This poses a challenge for learning and replaying slow sequences or rhythms with extended temporal gaps. 
Moreover, the time to the prediction of an upcoming sequence element, i.e., the onset of the dAP, as well as the time to the activation of the corresponding subset of neurons during autonomous replay is dictated by physiological parameters such as synaptic or membrane time constants, and independent of the inter-element intervals presented during learning. It is therefore unclear how the sTM model can learn (relative) time intervals or the duration of individual sequence elements. 
To represent the temporal structure of a sequence, the network requires a mechanism to encode time intervals between successive elements. The solution proposed here is to discretize time into elementary intervals shorter than the duration of a dendritic plateau potential and to construct longer intervals from a concatenation of these elementary intervals. In this study, this is realized by a delay line of sequentially activated neuronal assemblies. In the simplest case, these delay lines can be learned through repeated presentations of the same input.
The network thereby learns a sequence of distinct context-specific assemblies within the same minicolumn. Their sequential activation then encodes the duration of the corresponding element.
\par
\begin{figure}[ht]
  \centering
  \includegraphics[scale=\figscale]{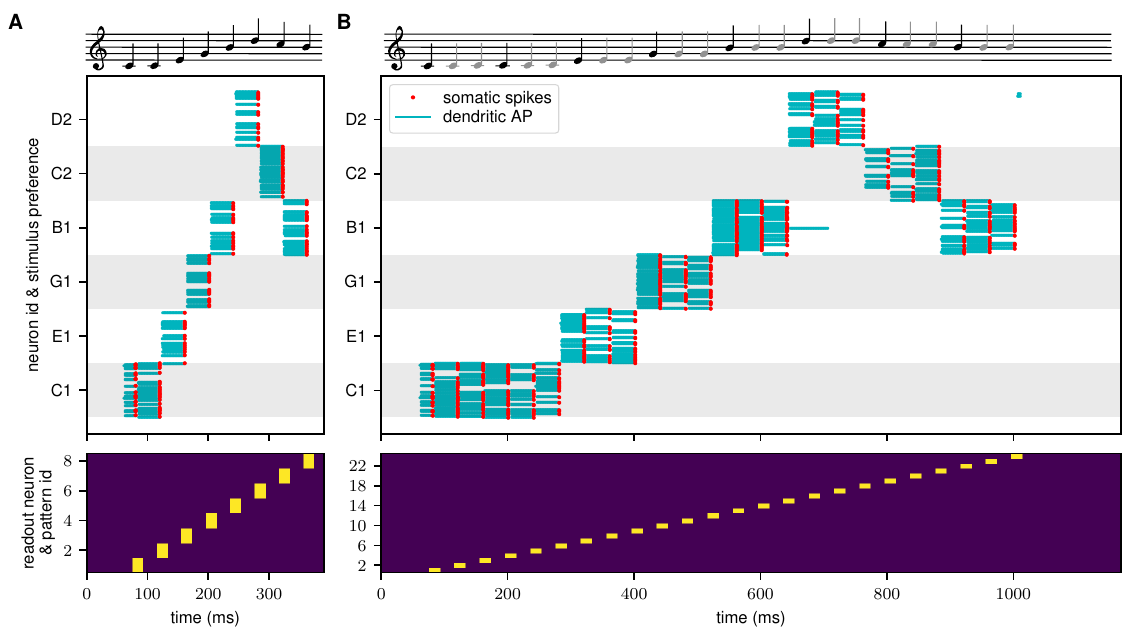}
  \caption{%
    \textbf{Encoding long intervals and rhythm through repetitive elements.}
    Network activity (middle row) after learning a melody (top row) presented at baseline speed (\textbf{A}) and a third of the baseline speed (\textbf{B}).
    Gray musical notes in the top row of panel B indicate repeated presentations of the preceding note, each with identical duration.
    Red dots in the middle row mark spikes, blue lines indicate dAP plateaus.
    The bottom row shows the readout neuron activity (violet: inactive, yellow: active), obtained by decoding the assembly identity from the spiking activity.
  }
  \label{fig:slow-sequences}
\end{figure}

Throughout this study, we consider a set of sTM networks, each trained on a single sequence
derived from a variation of a short melody (opening of ``Oh, Pretty Woman'' by Roy Orbison, 1964).
The base sequence \mbox{$s=\{\text{"C1","C1","E1","G1","B1","D2","C2","B1"}\}$} consists of $C=8$ elements drawn from a set of six musical notes (\cref{fig:slow-sequences}A).
Slower variants of this reference melody are constructed by repeating each note once or twice, yielding sequences of length $C=16$ and $C=24$.
During training, sequences are presented $500$ times, with a fixed inter-element interval of $\Delta T=40\,\ms$.
We refer to the inter-element interval during training as the encoding interval.
\par
\cref{fig:slow-sequences} illustrates our approach for two sTM networks trained on the reference melody at baseline speed ($C=8$) and on a slowed down version with a dilation factor of three ($C=24$), respectively. The temporal structure (``rhythm'') is represented by small populations of neurons, similar to a ``bar code'' that evolves over time. These bar codes provide an implicit representation of time elapsed since sequence onset, and enable a reliable decoding of sequence progression, i.e., the time since sequence start (see \cref{fig:slow-sequences} bottom panels).
\subsection{Controlling the speed of sequence replay}
\label{sec:replay_speed}
The sTM network can be configured into a ``replay mode'' by increasing the excitability of the excitatory neurons, such that the depolarization caused by a plateau potential alone is sufficient to elicit a spike.
In the original formulation of the sTM model \cite{Bouhadjar22_e1010233}, this is achieved by reducing the spike threshold of excitatory neurons.
From a dynamical perspective, this manipulation is equivalent to providing a constant background current to all excitatory neurons in the network.
In principle, this procedure permits a modulation of the replay speed by changing the magnitude of the background current.
In the following, we will however show that this mechanism offers only limited flexibility in controlling the speed at which activity propagates through the network.
An oscillatory background input, instead, provides a clock signal that can be used to flexibly and robustly regulate the speed of sequence replay.
\subsubsection{Modulating the replay speed with constant or oscillatory background inputs}
\label{sec:modulating_replay_speed}
\begin{figure}[ht]
  \centering
  \includegraphics[scale=\figscale]{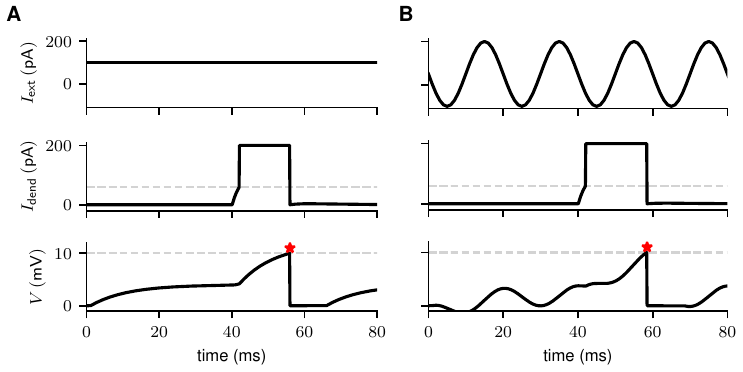}
  \caption{%
    \textbf{Illustration of the effect of different types of background activity on the membrane potential of an excitatory neuron in the sTM model.}
    The top row shows the external background current $I_\text{ext}$ applied to the neuron: a constant input of magnitude $\bar{I}=100\,\text{pA}$ (\textbf{A}) or an oscillatory input with frequency $f=50\,\text{Hz}$, amplitude $a=150\,\text{pA}$, and offset $\bar{I}=50\,\text{pA}$ (\textbf{B}). In both cases, the neuron additionally receives a predictive input to its dendrite at around $40\,\text{ms}$ (middle row). The resulting dendritic current $I_\text{dend}$ increases until it reaches a threshold $I_\text{dAP}$ (gray dashed line), triggering a plateau potential $I_\text{dAP}$. The superposition of background and dendritic input leads to a strong depolarization of the somatic membrane potential $V$ (bottom row) and a crossing of the spike threshold $V_\theta$ (gray dashed line). The neuron elicits a somatic spike (red star), which, in turn, triggers a reset of the dendritic current.
  }
    \label{fig:excitation-examples}
\end{figure}

A plateau potential generated by a dAP can directly trigger a somatic spike if it occurs during a phase of increased excitability, which may be caused by a sufficiently strong unspecific background input.
This background input could be a constant input (\cref{fig:excitation-examples}A), or a transient input, such as an oscillatory input (\cref{fig:excitation-examples}B).
As the background input is applied to all excitatory neurons in the sTM network, stable replay requires a proper choice of parameters: on the one hand, the background input must be strong enough to trigger a somatic spike in neurons that are in a predicted state.
On the other hand, it must be weak enough to leave unpredicted neurons below the firing threshold. 
Only the superposition of the plateau potential and the background input should elicit a spike.
Under these conditions, stable replay corresponds to the sequential activation of all neuronal assemblies encoding the sequence in the correct order, with no spurious activation of assemblies that do not belong to the replayed sequence.
The properties of the background input therefore determine whether stable replay of learned sequences is possible, and the speed at which learned sequences are replayed in response to a cue.
For a constant background input, the only critical parameter is its magnitude $\bar{I}$.
For a transient background input, not only the magnitude, but also the timing relative to the plateau potential matters.
Moreover, the duration of the high-excitability phase must be long enough to allow the membrane potential to reach the spike threshold.
For oscillatory inputs, sequence replay therefore depends on proper choices of the oscillation frequency $f$, the phase $\phi$, the amplitude $a$, and the offset $\bar{I}$.
In the following, we will compare sequence replay in sTM networks driven by constant or oscillatory background inputs, and investigate the range of parameters that permit a stable replay of learned sequences and how different parameter settings affect the replay speed.
\par
\begin{figure}[ht]
    \centering
    \includegraphics[scale=\figscale]{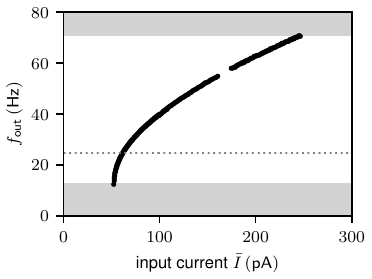}
    \caption{%
      \textbf{Modulation of replay speed by a constant background input.}
      Dependence of the replay speed $f_\text{out}$ on the magnitude $\bar{I}$ of a constant input current for the reference melody learned at baseline speed (\cref{fig:slow-sequences}A).
      The gray dotted line marks the encoding speed $f_\text{enc}=1/\Delta{}T=25\,\text{Hz}$. 
      Regions shaded in light gray correspond to parameter regimes where no replay is observed.}
    \label{fig:current-input}
\end{figure}
\begin{figure}[ht]
    \centering
    \includegraphics[scale=\figscale]{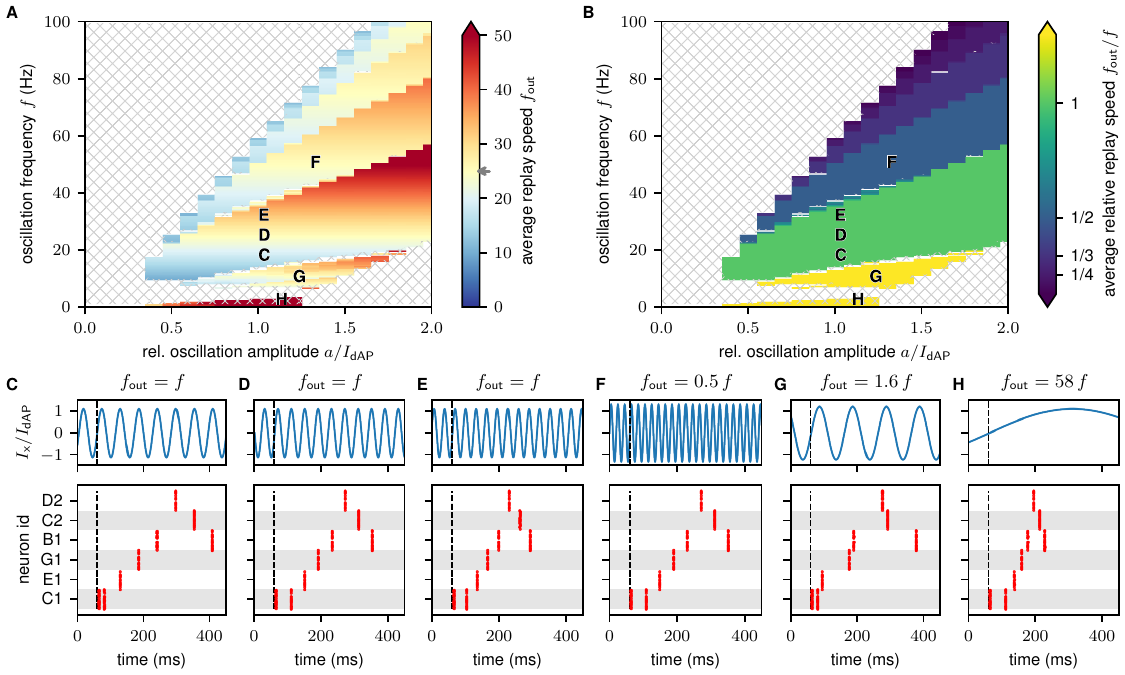}%
    \caption{%
      \textbf{Modulation of replay speed by oscillatory background input.}
      \textbf{A}: Dependence of the replay speed $f_\text{out}$ on the relative oscillation amplitude $a/I_\text{dAP}$ and the oscillation frequency $f$ during replay of the reference melody (\cref{fig:slow-sequences}A).
      The oscillation offset $\bar{I}$ is zero.
      The replay speed is averaged across eight different oscillation phases.
      The gray arrow in the colorbar marks the encoding speed $f_\text{enc}=1/\Delta{}T=25\,\text{Hz}$.
      Hatched areas represent parameter regions in which stable replay occurs for fewer than four of the eight phases.      
      \textbf{B}: Same as A, but for relative replay speed $f_\text{out}/f$.
      Green and blue bands correspond to regions where the replay speed is an integer fraction of the oscillation frequency, i.e., $f_\text{out}=f$, $f/2$, $f/3$, etc.
      Yellow regions indicate replay speeds exceeding the oscillation frequency, $f_\text{out}>f$.
      Bold letters in A and B mark the parameter settings illustrated in panels C--H.
      \textbf{C--H}: background oscillation (top) and corresponding replay activity (bottom) for different oscillation frequencies $f$ and amplitudes $a$: $f=18\,\text{Hz}$, $a/I_\text{dAP}=1.1$ (C), $f=25\,\text{Hz}$, $a/I_\text{dAP}=1.1$ (D), $f=32\,\text{Hz}$, $a/I_\text{dAP}=1.1$ (E), $f=50\,\text{Hz}$, $a/I_\text{dAP}=1.3$ (F), $f=10\,\text{Hz}$, $a/I_\text{dAP}=1.2$ (G), $f=1\,\text{Hz}$, $a/I_\text{dAP}=1.1$ (H). For all examples, the initial input triggers the replay at oscillation phase $\phi=0$ (vertical dashed lines).}
    \label{fig:oscillatory-input}
\end{figure}

For both constant and oscillatory background input, the network replays learned sequences over a range of speeds $f_\text{out}=10\ldots{}70\,\text{Hz}$ (\cref{fig:current-input} and \cref{fig:oscillatory-input}).
Here, the replay speed $f_\text{out}$ is defined as the inverse of the average inter-assembly interval during replay, computed from the last four activations of neuronal assemblies in the sequence.
The duration of the plateau potential $\tau_\text{dAP}=100\,\text{ms}$, which determines the maximum interval between consecutive assemblies, sets the lower bound of this range.
Spike transmission latencies for connections between excitatory neurons, rise times of dendritic synaptic currents, and membrane time constants of excitatory neurons  determine the upper bound.
The range of replay speeds is hence solely determined by neurophysiological parameters and the parameters of the background input.
The encoding speed $f_\text{enc}=1/\Delta{}T$, i.e., the speed at which assemblies are activated during learning (the inverse encoding interval) has no effect on the replay speed (provided it is not too small or too high for the sequence to be learned).
Through modulation of the background input parameters, sequence replay may be slower or faster than the encoding speed (cf.~gray dotted line in \cref{fig:current-input} and gray arrow in \cref{fig:oscillatory-input}A).
This mechanism therefore permits a modulation of the replay speed within certain bounds, without the need to relearn the sequence at the desired replay speed, and without the need to change intrinsic neuronal parameters. 
\par
Constant and oscillatory background inputs support similar replay speed ranges. However, the two types of inputs differ in how sensitively the replay speed depends on the input parameters. With a constant background current, slow replay requires fine tuning.
For replay speeds between $10\,\text{Hz}$ and $30\,\text{Hz}$, a tiny change in the input current has a strong effect (\cref{fig:current-input}).
Oscillatory background input, in contrast, provides a more robust and flexible control of the replay speed (\cref{fig:oscillatory-input}).
  Over wide parameter ranges, the replay speed scales linearly with the oscillation frequency.
  In the green band in \cref{fig:oscillatory-input}B, the replay speed matches the input frequency, $f_\text{out}=f$, such that successive oscillation cycles activate successive neuronal assemblies.
  Within this regime, the external oscillation acts as a ``clock signal'' determining the replay speed in a 1:1 manner. A broad range of replay speeds can be achieved by adjusting the oscillation frequency while keeping the amplitude fixed (vertical paths in \cref{fig:oscillatory-input}A,B and \cref{fig:oscillatory-input}C--E).
Unless the oscillation amplitude is adjusted accordingly, the low-pass filtering properties of the neuronal membrane lead to an attenuation of the membrane potential response as the oscillation frequency increases.
At some critical frequency, the superposition of the oscillatory input and the plateau potential may therefore no longer reach the spike threshold within a single oscillation cycle.
Beyond this point, the membrane may still reach the spike threshold by integrating the input over multiple oscillation cycles.
In this regime, successive assemblies are activated only at every second, third, or
higher-order oscillation cycle (blue bands in \cref{fig:oscillatory-input}B).
This transition to a regime where the replay speed is an integer fraction of the oscillation frequency is also observed when the oscillation amplitude is decreased while keeping the frequency fixed (horizontal paths in \cref{fig:oscillatory-input}A,B).
This effect allows the same input frequency to generate multiple replay speeds depending on the oscillation amplitude.
If the oscillation frequency is too high or the amplitude too low, replay is no longer possible (hatched regions in \cref{fig:oscillatory-input}A,B).
At low frequencies below $10\,\text{Hz}$, the replay speed exceeds the frequency $f$ of the oscillatory input.
Here, multiple assembly activations occur within a single oscillation peak, resulting in irregular replay (\cref{fig:oscillatory-input}G).
At very low frequencies $<5\,\text{Hz}$, the entire sequence may replay within a single oscillation peak (\cref{fig:oscillatory-input}H).
In this regime, sequence replay is fast and compressed, reminiscent of what is observed during slow-wave sleep in the hippocampus and in the neocortex \cite{Ji07,Euston07,Buzsaki15,Klinzing19}.
At these low frequencies, the oscillation period is longer than the duration of the plateau potential.
The predictive signal therefore decays before the next oscillation peak arrives. 
As a result, replay is effectively limited to a single oscillation peak, restricting the maximum sequence length \cite{Euston07}.
Moreover, replay success becomes sensitive to the oscillation phase at the time of replay initiation (see section \secref{sec:phase_invariance}).
\subsubsection{Range of accessible replay speeds}
\label{sec:limits_of_replay_speed}
\begin{figure}[ht]
    \centering
    \includegraphics[scale=\figscale]{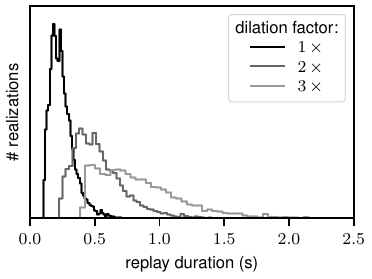}
    \caption{%
      \textbf{Range of replay durations achievable with oscillatory background input for sequences with different temporal dilation factors.}
      Distribution of replay durations across the ensemble of tested parameter combinations with
      $a\in\{0,0.1,\ldots,3\}\,I_\text{dAP}$, $f\in\{1,1.5,\ldots,100\}\,\text{Hz}$, $\phi\in\{0,1/4,\ldots,2\}\,\pi$, and $\bar{I}=0\,\text{pA}$ for a network trained on the reference melody with three different temporal dilation factors (see legend).
      The cases ``$1\times$'' and ``$3\times$'' are shown in \cref{fig:slow-sequences}A and B, respectively. 
    }
    \label{fig:durations}
\end{figure}

As shown in the previous section, global background input allows the sTM network to modulate replay speed over a range of approximately $10\,\text{Hz}$ to $70\,\text{Hz}$.
For our example sequence, the reference melody learned at baseline speed consisting of eight elements, this corresponds to a total replay duration, i.e., the time between the first and the last assembly activation, between about $90\,\text{ms}$ and $700\,\text{ms}$.
Within this range, the intrinsic neuronal and synaptic time constants limit the replay speed on the fast end.
On the slow end, it is limited by the duration of the plateau potentials.
Extending the replay duration beyond this temporal window requires modifying how the network represents the sequence.
As discussed in section \secref{sec:sequence_timing}, slower sequences can be learned by repeating elements during training such that each token is encoded by several successive neuronal assemblies within the same minicolumn.
During replay, the network must activate all of these assemblies sequentially, which elongates the path through the network and thereby increases total replay duration.
By varying the parameters of the oscillatory background input, a range of replay speeds can be achieved for each learned sequence representation (\cref{fig:durations}).
For the reference melody learned at baseline speed, the distribution of replay durations peaks at approximately $180\,\text{ms}$, corresponding to a replay speed of $f_\text{out}=40\,\text{Hz}$. 
When the system learns the same sequence at half or one third of baseline speed, longer replay durations become possible, reaching up to approximately $1.5\,\text{s}$ and $2.3\,\text{s}$, respectively.
This increase follows from the extended representational pathway in the network.
Lower learning speeds recruit more neuronal assemblies per sequence element, all of which must be traversed during replay.
If we assume that the inter-assembly interval during replay is constant across the sequence and given by $1/f_\text{out}$, the total replay duration can be approximated as $(C-1)/f_\text{out}$, where $C$ denotes the total number of assemblies used to encode the sequence.
Increasing $C$ hence leads to a shift of the entire  distribution of replay durations to larger values (\cref{fig:durations}).
Together, these examples illustrate that the total replay duration depends both on the background input driving replay and on the sequence speed during learning.
A sequence learned at some given speed can be replayed at a range of different speeds without relearning.
Extending this range, however, requires learning a new representation of the sequence.
\subsubsection{Phase invariance}
\label{sec:phase_invariance}
In natural settings, the timing of the external cue stimulus that triggers replay is unlikely to be locked to an intrinsic oscillation.
The phase of the oscillation at replay onset is therefore effectively random and can hardly be controlled \cite{Bouhadjar23_e1010989}.
To be biologically plausible, the mechanism proposed in this study for controlling the replay speed must hence be insensitive to a variability in the replay onset phase.
For many combinations of the oscillation frequency and amplitude, this is indeed the case (\cref{fig:oscillatory-input}A).
At low frequencies, however, replay driven by oscillatory background input becomes unreliable.
In this regime, successful replay depends strongly on the phase of the oscillation at the time of the cue.
If the cue occurs shortly before or during a trough in the background input, the predictive dendritic signal may decay before the next peak arrives, preventing the subsequent neuronal assembly from becoming active.
\par
\begin{figure}[ht]
  \includegraphics[scale=\figscale]{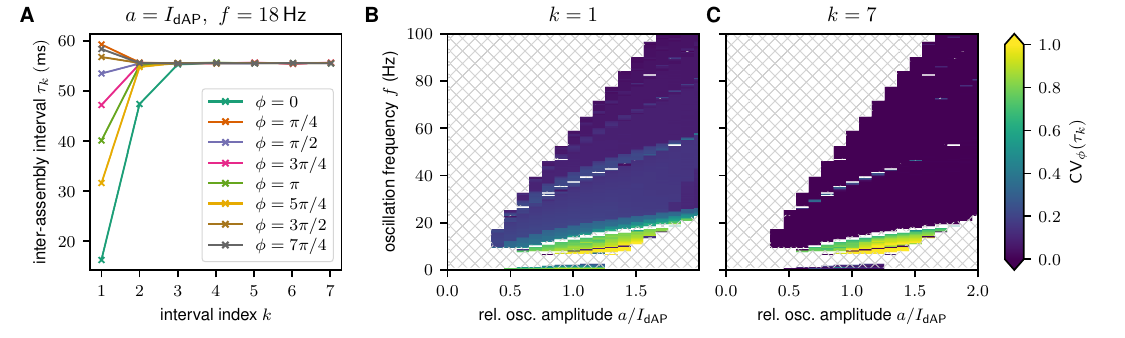}
    \caption{%
        \textbf{Phase invariance of replay speed.} 
        \textbf{A}: Dependence of the inter-assembly interval $\tau_\text{k}$ on interval index $k$ for eight different oscillation phases at replay onset (legend) of the reference melody at baseline speed (\cref{fig:slow-sequences}A). 
        \textbf{B,C}: Variability of replay timing, quantified by the coefficient of variation $\text{CV}_\phi(\tau_\text{k})$, of the first ($k=1$; B) and the last inter-assembly interval ($k=7$; C) as a function of oscillation frequency $f$ and relative amplitude $a/I_\text{dAP}$ with offset $\bar{I}=0\,\text{pA}$. 
        Hatched areas represent parameter regions where stable replay occurs for fewer than four of the eight phases. 
    }
    \label{fig:phase-invariance}
\end{figure}

Even if the predictive signal reliably leads to a response spike, the timing of this spike may differ for different onset phases.
Parameter regions that support stable replay may therefore exhibit variability in the replay speed.
Here, we investigate this phase dependence of the replay speed by analyzing the variability of the individual inter-assembly intervals during replay across different oscillation phases (\cref{fig:phase-invariance}).
For frequencies $f>20\,\text{Hz}$, the replay speed is largely invariant to the oscillation phase.
The first inter-assembly intervals after replay onset exhibit some small amount of variability, but this initial variability quickly disappears as replay progresses.
Only at low frequencies, the across-phase variability of the inter-assembly intervals is large and persists across the entire sequence.
\subsubsection{Robustness of replay in the presence of noise}
\label{sec:noise}
\begin{figure}[ht]
    \centering
    \includegraphics[scale=\figscale]{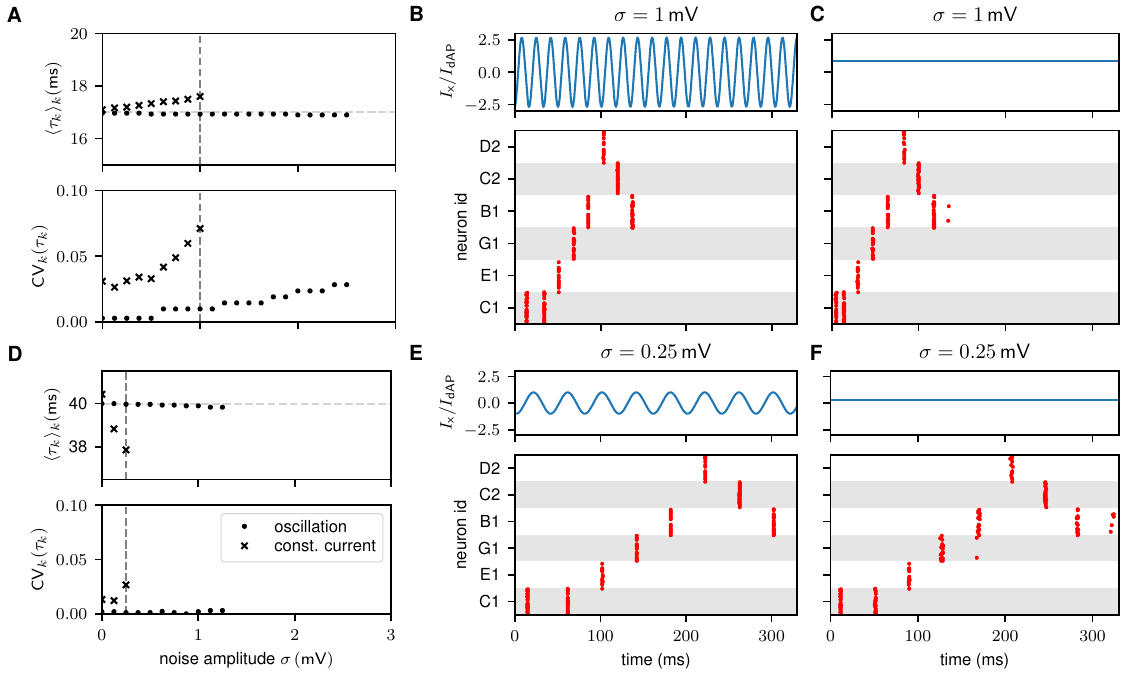}
    \caption{%
        \textbf{Stabilization of sequence replay by oscillatory input in the presence of noise.} 
        \textbf{A,D}: Dependence of the mean inter-assembly interval $\langle \tau_k \rangle_k$ (top panels) and the coefficient of variation $\text{CV}_k(\tau_k)$ of the inter-assembly intervals $\tau_k$ (bottom panels) on the amplitude $\sigma$ of a correlated background noise ($c=0.9$). 
        Mean inter-assembly intervals and variation coefficients are obtained from the last four elements ($k\in [4,7]$) of the sequence. 
        Crosses and circles represent data for constant and oscillatory background input, respectively (legend).   
        \textbf{B--F}: Background current (top) and spiking activity (bottom) for different configurations of background input:
        $f=59\,\text{Hz}$, $a/I_\text{dAP}=2.7$ (B),
        $\bar{I}/I_\text{dAP}=0.885$ (C),      
        $f=25\,\text{Hz}$, $a/I_\text{dAP}=1$ (E),
        $\bar{I}/I_\text{dAP}=0.305$ (F)
        with $\phi=0$ and $\bar{I}=0\,\text{pA}$. 
        Examples are shown at noise amplitudes $\sigma$ corresponding to the highest noise levels at which replay remains stable for both constant and oscillatory background input (see vertical dashed lines in A,D).
        Top and bottom rows correspond to parameter combinations resulting in a faster and a slower replay, respectively ($\langle\tau_k\rangle_k = 17\,\text{ms}$ and $40\,\text{ms}$ for $\sigma=0\,\text{mV}$).
        Data shown here are obtained from a single noise realization.
      }
      \label{fig:noise}
\end{figure}

In the present study, we model the background input as a constant or an oscillatory current injected into the excitatory neurons of the sTM network.
In the biological system, this background input is the result of synaptic inputs from presynaptic sources firing in a tonic or oscillatory manner.
The net background currents are therefore subject to fluctuations due to the variability in the presynaptic spike times.
Here, we investigate the robustness of sequence replay in the presence of such noise for varying replay speeds and for both constant and oscillatory background input (\cref{fig:noise}).
\par
To account for the close spatial proximity and shared input statistics of neurons within a minicolumn, we model background noise as an additive shot-noise process (low-pass filtered Poissonian spike trains) with a high correlation coefficient $c=0.9$ (see \cref{eq:shot_noise} in \secref{sec:methods}).
Thus, neurons within a minicolumn receive $90\%$ of their noise from a shared source and the remaining $10\%$ from independent sources.
Neurons in different minicolumns receive independent noise.
We parameterize the noise amplitude by the standard deviation $\sigma$ of the resulting fluctuations in the neuronal membrane potential.
\par
In the presence of noise, some neurons in an activated assembly reach the spike threshold earlier than others.
Hence, increasing the noise amplitude $\sigma$ leads to a dispersion of spike times within an assembly.
This desynchronization eventually causes failures in triggering dAPs in the next assembly, and ultimately leads to a breakdown of replay.
The noise induced desynchronization can be counteracted to some extent by increasing the level of noise correlation within each minicolumn, which keeps the spike times of neurons within the same assembly aligned even at moderate noise amplitudes (see \cite{Bouhadjar23_e1010989}).
However, some fraction of synaptic input noise is likely to be uncorrelated across neurons (here, $10\%$), such that the desynchronizing effect of noise cannot fully be avoided.
For transient (e.g., oscillatory) background input, the desynchronizing effect of noise is reduced by the mechanisms described in \cite{Mainen95_1503,Mohns99,Goedeke08_015007}: transient inputs lead to a faster threshold crossing, such that there is less time for noise to accumulate.
As a result, the spike times of neurons in the same assembly are less dispersed.
Overall, we observe that sequence replay is more robust to noise when driven by oscillatory background input, as compared to constant background input.
Oscillatory background input permits larger (and biologically realistic) noise amplitudes (\cref{fig:noise}A,D, top panels).
The range of tolerable noise amplitudes is larger for faster replay, because the spike threshold is approached more quickly, which, in turn, synchronizes assembly spikes.
This effect is observed for both constant and oscillatory background input.
Within the range of tolerable noise amplitudes, the variability of the inter-assembly intervals across sequence elements increases more rapidly with $\sigma$ if the background is constant, as compared to oscillatory background input (\cref{fig:noise}A,D, bottom panels).
In other words, with oscillatory background input, the replay speed remains more stable across sequence elements.
Similarly, in the presence of oscillatory background input, the replay speed remains constant when increasing the noise level, up to the point where replay breaks down (\cref{fig:noise}A,D, top panels).
With constant background input, in contrast, the replay speed systematically changes when increasing the noise level.
\subsection{Testable predictions}
\label{sec:testable_predictions}
The mechanisms proposed in this work give rise to several experimentally testable predictions .
\subsubsection{Working memory with finite duration through stereotyped sequential activation of sparse assemblies}

The proposed mechanism represents extended temporal intervals by repeated activation of stimulus-specific, sparse neuronal subsets within the same minicolumn (Fig.~\ref{fig:slow-sequences}).
This suggests that working memory for temporally extended stimuli relies on a progression through a finite sequence of distinct assemblies.
Experimentally, this predicts that during maintenance of rhythmic or temporally extended stimuli, neural recordings should reveal sequential activation of stimulus-specific subpopulations even when the external input is constant.
The duration of a remembered item should correlate with the number of assembly activations rather than with the persistence of a single active ensemble.
High-resolution recordings should therefore reveal a ``bar code''-like temporal evolution of sparse population activity.
\par
This representation further implies a cost for encoding slow sequences.
Longer intervals require more assemblies and increased spiking activity, making them more demanding to learn.
Accordingly, learning slower sequences should require more training trials than faster ones.
Consistent with this idea, human psychophysics shows improved temporal precision at faster speeds, while slower sequences exhibit increased variability \cite{Hardy18}.
\subsubsection{Replay speed is controlled by global oscillatory dynamics}
According to the mechanism proposed in this study, the speed of sequence replay in sTM networks scales with the frequency of a global oscillatory input (Fig.~\ref{fig:oscillatory-input}).
This predicts that the speed of replay in the hippocampus and in the neocortex correlates with the dominant frequency of ongoing oscillations measurable in EEG or LFP signals. Moreover, during low-frequency oscillations ($1$--$4\,\text{Hz}$), the number of sequence elements that can be replayed within a single cycle should be limited.
Consistent with this, replay in bats during sleep is temporally compressed but fragmented, suggesting constraints on replay duration \cite{Eliav25}.
In addition, replay onset during slow-wave sleep should be phase-locked to the underlying oscillation, as successful initiation requires a favorable phase.
\subsubsection{Oscillation frequency determines the perceived duration of time intervals}
The proposed mechanism predicts a direct link between the frequency of global background oscillations and the subjective estimation of time.
In our framework, elapsed time is encoded by the sequential progression of activity through an assembly chain, and the replay speed scales with oscillation frequency.
When replay proceeds faster than the speed at which the sequence was learned, activity traverses the assembly chain more rapidly, causing the network to reach assemblies associated with later sequence elements sooner.
This corresponds to an overestimation of elapsed time: the internal representation signals that more time has passed than has actually elapsed.
Conversely, slower replay driven by lower oscillation frequencies should lead to an underestimation of elapsed time.
These predictions hold within each of the distinct bands identified in \cref{fig:oscillatory-input}B, where replay speed scales linearly with the oscillation frequency.
These predictions are consistent with experimental findings showing that pharmacological and physiological manipulations affecting the speed of neural dynamics systematically bias time estimation.
Increased dopamine levels, which have been associated with accelerated neural dynamics, lead to a prolongation of estimated time intervals (see \cite{Hass16} and references therein).
Cooling of the brain during duration estimation tasks in rats causes a slowing of neuronal dynamics and a corresponding underestimation of elapsed time \cite{Monteiro23}. 

\subsubsection{Replay speed correlates with sequence speed during learning}
The temporal range over which a sequence can be replayed depends on how it was encoded during learning (Fig.~\ref{fig:durations}).
Sequences learned at faster speeds have shorter representational chains and therefore more restricted replay ranges.
This leads to two predictions: (i) replay duration should correlate with the speed at which the sequence was learned, and (ii) replay at speeds far outside the learning regime may require relearning at a different speed.
These predictions can be tested in psychophysics experiments.
\subsubsection{Rapid fading of across-phase interval variability}
  In the model presented here, the time intervals between the first elements of a replayed sequence often exhibit some degree of variability across different oscillation phases.
  Typically, this initial interval variability quickly disappears as the replay progresses, such that the intervals between later sequence elements are largely invariant to the oscillation phase at replay onset (Fig.~\ref{fig:phase-invariance}).
  This fading of across-trial interval variability during replay should be observable in neural recordings of replay events, and could be tested in psychophysics experiments.
  Indeed, psychophysics experiments investigating the precision of time estimation are in line with our prediction:
  in general, the across-trial variability (standard deviation) of estimated time grows linearly with the length of the time interval to be estimated.
  This finding is known as ``Weber's law'' \cite{Gibbon77_279,Hass16}, and can be explained by an accumulation of noise across successive assembly transitions during replay.
  At short and at long time intervals, however, the variability of time estimation is often disproportionately large, leading to deviations from Weber's law \cite{Bizo06_201}.
  The larger variability at short intervals could be explained by the phase-dependent variability of the initial replay intervals, which dominates when the assembly chain encoding the estimated time interval is short. 
\subsubsection{Oscillations enhance robustness against noise}
Oscillatory input stabilizes replay by periodically aligning neuronal excitability (Fig.~\ref{fig:noise}).
Replay during strong coherent oscillations should therefore be more temporally precise and less susceptible to noise, whereas low-amplitude oscillatory states should exhibit higher variability and failure rates.
This prediction could be tested using analyses similar to those in \cite{Denker11_2681}.
The prediction is consistent with experimental findings showing that temporally structured inputs increase spike-time reliability  \cite{Mainen95_1503} and that slow cortical rhythms provide a temporal reference frame that stabilizes neural representations \cite{Kayser09}.
The noise-reducing effect of oscillations should moreover have an influence on the variability of time estimation: the slope of Weber's law (see above) should depend on the characteristics of background oscillations.
\subsubsection{Timing sensitive replay relies on stable oscillations}
Oscillations in cortical activity are often transient and exhibit high levels of variability in frequency, amplitude, and phase \cite{Burns11_9658,Xing12_13873}.
The mechanism proposed here suggests that during tasks where timing matters, the global oscillations serving as a clock signal must be stable over the timescale of the sequence to be replayed (in the relevant cortex area).
\section{Discussion}
\label{sec:discussion}
This study extends the spiking Temporal Memory (sTM) model to address two key aspects of sequence processing: the learning and the representation of sequence element durations or time intervals between sequence elements, and the control of the speed of sequence replay after learning. We show that the temporal structure of sequences can be encoded in the sTM model by a sequential activation of context-specific neuronal assemblies, such that elapsed time corresponds to a progression along an assembly chain. This permits a discretized representation of time that preserves sparse, context-dependent coding, and supports an unsupervised learning of sequences with complex temporal structure via local synaptic plasticity mechanisms. In addition, we demonstrate that synaptic background input can control the speed of autonomous sequence replay in response to a cue signal. While constant background input requires fine tuning, oscillatory background input provides a robust clock signal by which the replay speed can be flexibly modulated. This establishes a functional separation between the representation of the temporal structure of a sequence, i.e., the sequence of relative time intervals between consecutive sequence elements (encoded in the network connectivity), and of the temporal scale, i.e., the overall temporal compression or dilation of the sequence, during autonomous replay (controlled by global background input). Our results reveal constraints on the replay dynamics imposed by neurophysiological parameters: spike transmission delays, membrane time constants, and synaptic time constants limit the maximum replay speed, whereas the duration of plateau potentials caused by dendritic action potentials (dAPs) sets a lower bound. Extending the range of replay speeds beyond these limits requires modifying the learned representations by adding or removing assemblies in the chain of neuronal populations representing the sequence. As a result, the range of achievable replay speeds depends on the temporal structure of the sequence present during learning. Together, we suggest that both the contents and the temporal structure of a learned sequence are co-represented by the same chain of assemblies formed by small, context-specific groups of cortical neurons. 
These assembly chains constitute the neuronal substrate for a well timed prediction of upcoming events or detection of anomalies, as well as for the autonomous replay of temporally structured sequences in response to a cue.
The speed of sequence replay, i.e., the speed of activity propagation along these assembly chains, can be modulated within some range by global oscillatory activity.
Based on these ideas and results, we formulate a set of testable predictions for an experimental validation (see section \secref{sec:testable_predictions}).
\par
The central outcome of this study is a reconciliation of the sTM model with the ability to learn and control timing in sequential data.
The sTM model combines a number of features that are not commonly found together in other models of sequence learning and processing: it learns sequences in a continual unsupervised manner via local plasticity rules, it accounts for context (history) dependence in sequential data with a high context depth, it exhibits high storage capacity and low energy demands due to its ultra-sparse coding, it supports different operation modes (prediction, anomaly detection, replay), and it is biologically plausible in terms of its neuronal and synaptic mechanisms.
This study extends this list of features by showing that the sTM model can also learn and represent complex temporal structure in sequential data, and that it permits a flexible and robust control of the speed of sequence replay after learning.
\par
The timing mechanisms employed in this study are not entirely new.
There exists a large number of models of time encoding and perception in computational neuroscience.
An overview is provided in \cite{Hass16}.
This study classifies existing models of time encoding and perception into four categories: 1) ramping activity models, 2) delay-line models, 3) state-dependent timing models, and 4) neural oscillator models.
The model presented in our study falls into the second category: a delay-line model where time intervals are encoded by the position of activity along a sequence of neuronal groups.
While each individual sequence in the sTM model is represented by such a chain of neuronal groups, the sTM network is still a highly recurrent network.
The same neuron can, for example, participate multiple times in the same or in different sequences.
Assembly chains representing different sequences may overlap.
Each individual minicolumn may be activated repeatedly when encoding longer time intervals, which corresponds to a high degree of feedback within minicolumns.
One of the key advantages of delay-line models is their ability to store temporal structure within the connectivity of the network.
The representation of time information is therefore very robust and does not rely on a fine tuning of parameters such as synaptic weights or time constants \cite{Roxin04,Vogels05_10786,Lundqvist06_253,Gavornik11_501,Kriener2014_136}.
A drawback is that longer time intervals require more neurons.
As shown in \cite{Hass-2008_449}, delay-line models can explain psychophysical results obtained from humans such as the Weber law and deviations from it at short and long time intervals (see also section \secref{sec:testable_predictions}).
Delay-line models are therefore plausible candidates for the storage of temporal information in biological neuronal circuits.
The model proposed in our study extends the basic delay-line model by combining it with a neural oscillator model, which provides a mechanism for controlling the replay speed after learning without the need to relearn the sequence representation.
Moreover, our model permits the storage of timing and contents information within the same neuronal circuit.
It does not require a separate timing network (as, for example, in \cite{Maes20_e1007606,Calderon22,Liang20}). 
\par
A number of earlier studies propose that sequences can be represented by transitions between attractor states in recurrent neuronal networks, i.e., in networks where individual sequence elements are stored within clusters of neurons in an auto-associative manner, and where additional hetero-associative connections are used to implement transitions between these attractors (see, e.g., \cite{Palm82,Horn89_1036,Wennekers09_429,Lansner13_e73776, Martinez19_e0220161, Pereira20_97, Cone21_e63751}).
One of the challenges in these models is to explain how transitions between attractor states can be initiated and how the timing of these transitions can be controlled.
\cite{Wennekers09_429} suggest that these transition could be triggered by well timed, unspecific external signals (see also ``pump of thought'' in \cite{Braitenberg84}).
This leaves open the question of how the timing of these external signals is controlled.
In our model, we also employ an unspecific external signal to trigger transitions between sequence elements during replay.
Here, however, the ongoing oscillatory background inputs are only used to control the speed of a learned sequence.
The relative time intervals between sequence elements are stored within the network structure (number of assemblies in the delay line).
Other studies employ intrinsic dynamical mechanisms, such as short-term synaptic plasticity, adaptation, or transient network dynamics, to trigger transitions between attractors (see, e.g., \cite{Lansner13_e73776, Martinez19_e0220161, Pereira20_97, Cone21_e63751}).
With this approach, however, a precise control of the timing typically requires a fine tuning of parameters such as synaptic weights or time constants \cite{Roxin04,Vogels05_10786,Lundqvist06_253,Gavornik11_501,Kriener2014_136}.
Storing the temporal structure of a sequence in the network connectivity instead, as in the sTM model, allows for a much more robust representation of time that does not require fine tuning of parameters.
\par
The model proposed in this study suggests a trade-off between the number of neurons required to represent a given time interval and the flexibility in modulating the replay speed.
The total time interval represented by an assembly chain is determined by the number of assemblies and the encoding interval, i.e., the time interval between successive assemblies during training.
A given time interval could be realized by a chain composed of many assemblies with short encoding intervals, or by a short chain with longer encoding intervals.
Shorter chains require fewer neurons.
However, if the encoding interval is large and close to the maximum possible interval (the duration of the dAP plateau potential), the sequence cannot be slowed down during replay.
If, in contrast, the encoding interval is close to its minimum (determined by synaptic and membrane time constants), the sequence cannot be sped up during replay (and requires more neurons).
Maximum flexibility in the replay speed modulation is therefore achieved if the encoding interval is in the intermediate range, at about one half of the duration of the dAP plateau potentials.
A related trade-off has been discussed in the context of other delay-line models \cite{Hass08_449}.
\par
For the variant of the model presented in this study, long time intervals between successive sequence elements exceeding the duration of a plateau potential are learned by repeated activation of the minicolumn corresponding to the previous sequence element, which leads to the recruitment of additional assemblies within the same minicolumn.
As an alternative, the temporal gap between consecutive sequence elements could be filled by recruiting assemblies in other minicolumns encoding ``silence''.
A combination of these two encoding strategies may allow the model to capture more complex temporal structures, such as sequences containing both sustained elements and pauses.
\par
Representing time through sequential state transitions naturally relates to time cells observed in hippocampus, striatum, and cortex \cite{Eichenbaum14_732,Tiganj17_5663}.
In both cases, elapsed time is encoded by a progression of neuronal activity patterns.
In the present framework, however, longer intervals require more transitions, increasing resource usage and allowing noise to accumulate \cite{Hass08_449}.
This may contribute to a reduced temporal precision for longer durations, consistent with experimental observations \cite{Tiganj17_5663,Cao22}.
In section \secref{sec:noise}, we show that constant and oscillatory background inputs differ in how they stabilize replay dynamics in the presence of noise.
This accumulation of variability across successive transitions has been shown to give rise to Weber's law, i.e., a proportional increase of timing variability with interval duration \cite{Hass08_449}.
As discussed in section \secref{sec:testable_predictions}, our framework suggests that this scaling may depend on the properties of the background input.
A direction for future work is therefore to investigate how different background input regimes modulate this scaling.
\par
Oscillations in the present model act as a global clock signal controlling sequence progression.
In the neuroscientific literature, oscillations are associated with additional functions, including communication through coherence \cite{Fries05_474}, probabilistic inference and exploration \cite{korcsak2022cortical,Bouhadjar23_e1010989}, and phase-based coding of spike timing \cite{Montemurro08,Kayser12}.
Rather than being limited to a single role, oscillations may simultaneously structure neural dynamics across multiple timescales.
Understanding how such functions interact with the clocking mechanism proposed here remains an important direction for future work.
\par
The model presented in this study provides a potential link to replay phenomena observed during sleep.
Replay events in hippocampus and cortex are often temporally compressed and occur within transient windows of synchronized network activity \cite{Ji07,Euston07,Carr12,Buzsaki15,Klinzing19}.
In our framework, similar fast and phase-dependent replay emerges under low-frequency oscillatory input, where successful propagation depends on the timing of initiation relative to the oscillation cycle.
This regime limits the number of sequence elements that can be traversed and may lead to fragmented replay, consistent with experimental observations \cite{Eliav25}.
At the same time, temporal compression during replay may facilitate synaptic plasticity by bringing successive assemblies closer together in time, potentially enabling the formation of shortcut connections between non-adjacent elements \cite{Gupta10,Pfeiffer13}.
\par
Overall, our results show that temporal structure can be encoded by sequential activation of sparse neuronal assemblies, while global oscillatory input provides a flexible and robust mechanism for controlling the speed of sequence replay.
This suggests that temporal structure in neural computation may emerge from interactions between sequential population dynamics and global oscillatory states.
\section{Acknowledgments}
\begin{sloppypar}
This project was supported by the Federal Ministry of Education and Research (NeuroSys grant no.~03ZU1106CB, 03ZU2106CB; NEUROTEC-II grant no.~16ME0398K, 16ME0399), the German Research Foundation grant no.~368482240/GRK2416, the Joint Lab ``Supercomputing and Modeling for the Human Brain'' (SMHB), the European Union’s Horizon Europe Programme under the Specific Grant Agreement No. 101147319 (EBRAINS 2.0 Project), and Joint Lab HiRSE the Helmholtz Platform for Research Software Engineering---an innovation pool project of the Helmholtz Association.
\end{sloppypar}
\section{Author contributions}
\textbf{Conceptualization:} Melissa Lober, Markus Diesmann, Tom Tetzlaff.\\
\textbf{Data curation:} Melissa Lober.\\
\textbf{Formal analysis:} Melissa Lober.\\
\textbf{Funding aquisition:} Markus Diesmann.\\
\textbf{Investigation:} Melissa Lober, Tom Tetzlaff.\\
\textbf{Methodology:} Melissa Lober, Younes Bouhadjar, Tom Tetzlaff.\\
\textbf{Project administration:} Melissa Lober, Markus Diesmann, Tom Tetzlaff.\\
\textbf{Software:} Melissa Lober, Younes Bouhadjar, Markus Diesmann, Tom Tetzlaff.\\
\textbf{Supervision:} Markus Diesmann, Tom Tetzlaff.\\
\textbf{Visualization:} Melissa Lober.\\
\textbf{Writing --- original draft:} Melissa Lober.\\
\textbf{Writing --- reviewing \& editing:} Melissa Lober, Younes Bouhadjar, Markus Diesmann, Tom Tetzlaff.
\section{Data availability statement}
The source code and the simulation data required to reproduce the results of this study are openly available \cite{Lober26_20290700}.
\clearpage
\section{Methods}
\label{sec:methods}
\subsection{Detailed description of the network model}
\renewcommand{\labelitemi}{$\bullet$}
\renewcommand{\labelitemii}{$\circ$}
\renewcommand{\labelitemiii}{{\tiny$\blacksquare$}}
\renewcommand{\labelitemiv}{{\tiny$\square$}}
\begin{table}[H]
\renewcommand{\arraystretch}{1.1}
\begin{tabular}{|@{\hspace*{1mm}}p{3cm}@{}|@{\hspace*{1mm}}p{12cm}|}
\hline 
\multicolumn{2}{|>{\color{white}\columncolor{black}}c|}{\textbf{Summary}}\\
\hline
\textbf{Populations} &  excitatory ($\Epop$), inhibitory ($\Ipop$), bottom-up ($\Xpop$), and top-down ($\Tpop$)  \\
\hline 
\textbf{Connectivity} &
\begin{itemize}
    \item sparse random connectivity between excitatory neurons (plastic)
    \item local recurrent connectivity between excitatory and inhibitory neurons (static)
\end{itemize}
\\
\hline
\textbf{Neuron model} & 
\begin{itemize}
\item excitatory neurons: leaky integrate-and-fire (LIF) with nonlinear input integration (dendritic action potentials)     
\item inhibitory neurons: leaky integrate-and-fire (LIF)
\end{itemize}
\\
\hline 
\textbf{Synapse model } & exponential or alpha-shaped postsynaptic currents (PSCs)  \\
\hline 
\textbf{Plasticity } &  spike-timing dependent structural plasticity and weight decay in excitatory to excitatory connections
\\
\hline 
\textbf{Input} & bottom-up and top-down spike sources and background currents, connected to excitatory neurons \\
\hline
\end{tabular}
\caption{Summary of the network model. Parameter values are given in Table \cref{tab:Model-parameters}.}
\label{tab:Model-description-summary}
\end{table}
\begin{table}[H]
\begin{tabular}{|@{\hspace*{1mm}}p{3cm}@{}|@{\hspace*{1mm}}p{10.95cm}@{}|@{\hspace*{1mm}}p{0.95cm}|}  
\hline 
\multicolumn{3}{|>{\color{white}\columncolor{black}}c|}{\textbf{Populations}}\\
  \hline 
  \textbf{Name} & \textbf{Elements} & \textbf{Size}\\
\hline
 $\Epop_k$ & excitatory neurons in subpopulation $k$, \mbox{$\Epop_k\cap\Epop_l=\emptyset\ (\forall{}k\ne{}l\in[1,M])$} & $n_\exc$ \\
  \hline
 $\Epop=\bigcup_{k=1}^M\Epop_k$ & excitatory (E) neurons  & $N_\exc$\\
  \hline
  $\Ipop_k$ & inhibitory neurons in subpopulation $k$, \mbox{$\Ipop_k\cap\Ipop_l=\emptyset\ (\forall{}k\ne{}l\in[1,M])$} & $n_\inh$ \\
  \hline
  $\Ipop=\bigcup_{k=1}^M\Ipop_k$ & inhibitory (I) neurons & $N_\inh$\\
  \hline
  $\Xpop_k$ & spike generator for bottom-up input to subpopulation $k$, \mbox{$\Xpop_k\cap\Xpop_l=\emptyset\ (\forall{}k\ne{}l\in[1,M])$} & $1$ \\
  \hline
 $\Xpop=\bigcup_{k=1}^M\Xpop_k$ & bottom-up spike generators  & $M$\\
  \hline
  $\Tpop_i$ & spike generator triggering top-down prediction of sequence $i$, \mbox{$\Tpop_i\cap\Tpop_j=\emptyset\ (\forall{}i\ne{}j\in[1,S])$} & $1$ \\
  \hline
 $\Tpop=\bigcup_{i=1}^S\Tpop_i$ & spike generators triggering top-down predictions  & $S$\\
  \hline
   $\pi_{i,1}$ & excitatory neurons representing the first element of sequence $i$\newline ("assembly"; \mbox{$\forall{}i\in[1,S]$}) & $\rho$\\
  \hline
\end{tabular}
\caption{Description of the populations. Parameter values are given in \cref{tab:Model-parameters}. }
\label{tab:Model-description-populations}
\end{table}
\begin{table}[!ht]
  \renewcommand{\arraystretch}{1.2}
  \small
\begin{tabular}{|@{\hspace*{1mm}}p{1.85cm}@{}|@{\hspace*{1mm}}p{1.85cm}@{}|@{\hspace*{1mm}}p{11.2cm}|}
\hline 
\multicolumn{3}{|>{\color{white}\columncolor{black}}c|}{\textbf{Connectivity}}\\
\hline 
\textbf{Source} & \textbf{Target} & \textbf{Pattern}\\
\hline 
  $\Epop$  & $\Epop$ &
                       random;
                       fixed in-degrees $K_i=K_\EE$, delays $d_{ij}=d_{\EE}$, and synaptic time constants $\tau_{ij}=\tau_{\EE}$, 
                       plastic synaptic weights $J_{ij}$ 
                       ($\forall{}i\in{\Epop},\,\forall{}j\in{\Epop}$; ``$\EE$ connections'');  no self-connections (``autapses''), no multiple connections (``multapses'') \\
\hline 
  $\Epop_k$  & $\Ipop_k$ & all-to-all,
                       fixed delays $d_{ij}=d_{\IE}$, synaptic time constants $\tau_{ij}=\tau_{\IE}$, and weights $J_{ij}=J_\IE$
                       ($\forall{}k\in[1,M]$, $\forall{}i\in\mathcal\Ipop_k,\,\forall{}j\in{}\Epop_k$; ``$\IE$ connections'') \\
\hline 
  $\Ipop$ & $\Epop$ & all-to-all;
                      fixed delays $d_{ij}=d_{\EI}$, synaptic time constants $\tau_{ij}=\tau_{\EI}$, and weights $J_{ij}=J_\EI$
                      ($\forall{}i\in\Epop,\,\forall{}j\in{}\Ipop$; ``$\EI$ connections'') \\
\hline
  $\Xpop_k$ & $\Epop_k$ & all-to-all, fixed delays $d_{ij}=d_{\EX}$, synaptic time constants $\tau_{ij}=\tau_{\EX}$, and weights $J_{ij}=J_\EX$
                       ($\forall{}k\in[1,M]$, $\forall{}i\in\mathcal\Epop_k,\,\forall{}j\in{}\Xpop_k$; ``$\EX$ connections'')\\
\hline
  $\Tpop_h$ & $\pi_{h,1}$ & all-to-all, fixed delays $d_{ij}=d_{\ET}$, synaptic time constants $\tau_{ij}=\tau_{\ET}$, and weights $J_{ij}=J_\ET$
                       ($\forall{}h\in[1,S]$, $\forall{}i\in\mathcal\pi_{h,1},\,\forall{}j\in{}\Tpop_h$; ``$\ET$ connections''); no multiple connections (``multapses'')\\
\hline
\multicolumn{3}{|>{}l|}{all unmentioned connections are absent}\\
\hline
\end{tabular}
\caption{Description of the connectivity. Parameter values are given in \cref{tab:Model-parameters}.}
\label{tab:Model-description-connectivity}
\end{table}
\begin{table}[ht!]
\begin{tabular}{|@{\hspace*{1mm}}p{3cm}@{}|@{\hspace*{1mm}}p{12cm}|}
  \hline 
  \multicolumn{2}{|>{\color{white}\columncolor{black}}c|}{\textbf{Neuron}}\\
  \hline
  \textbf{Type} & leaky integrate-and-fire (LIF) dynamics \\
  \hline
    \textbf{Description} & dynamics of membrane potential $V_{i}(t)$  and spiking activity $\xi_i(t)$ of neuron $i$:                
            \begin{itemize}
              \item emission of the $k$th spike of neuron $i$ at time $t_{i}^{k}$ if
                \begin{equation}
                  V_{i}(t_{i}^{k})\geq\theta_i 
                \end{equation}
                 with somatic spike threshold $\theta_i$
              \item spike train: $\xi_i(t)=\sum_k\delta(t-t_i^k)$
              \item reset and refractoriness:
                \begin{equation*}
                  V_{i}(t)=\Vreset
                    \quad \forall{}k,\ \forall t \in \left(t_{i}^{k},\,t_{i}^{k}+\tau_{\text{ref},i}\right]
                \end{equation*}
                with refractory time $\tau_{\text{ref},i}$ and reset potential $\Vreset$
              \item subthreshold dynamics:
                \begin{equation}
                  \label{eq:lif}
                  \tau_{\text{m},i}\dot{V}_i(t)=-V_i(t)+R_{\text{m},i} I_i(t)
                \end{equation}
                with membrane resistance $R_{\text{m},i}=\dfrac{\tau_{\text{m},i}}{C_{\text{m},i}}$, membrane time constant $\tau_{\text{m},i}$, and total synaptic input current $I_i(t)$ (see \cref{tab:Model-description-synapse})
              \item excitatory neurons: $\tau_{\text{m},i}=\tau_\text{m,E}$, $C_{\text{m},i}=C_\text{m}$, $\theta_i=\theta_\text{E}$, $\tau_{\text{ref},i}=\tau_\text{ref,E}$ ($\forall i\in\Epop$)
              \item inhibitory neurons: $\tau_{\text{m},i}=\tau_{\text{m},I}$, $C_{\text{m},i}=C_\text{m}$, $\theta_i=\theta_\text{I}$, $\tau_{\text{ref},i}=\tau_\text{ref,I}$ ($\forall i\in\Ipop$)       
   
            \end{itemize}\\
              \hline 
\end{tabular}
\caption{Description of the neuron model. Parameter values are given in \cref{tab:Model-parameters}.}
\label{tab:Model-description-neuron}
\end{table}
\begin{table}[ht!]
  \small
  \begin{tabular}{|@{\hspace*{1mm}}p{3cm}@{}|@{\hspace*{1mm}}p{12cm}|}
  \hline
  \multicolumn{2}{|>{\color{white}\columncolor{black}}c|}{\textbf{Synapse}}\\
  \hline
  \textbf{Type} & continuous, exponential, or alpha-shaped postsynaptic currents (PSCs) \\
  \hline
  \textbf{Description} &                 
    \begin{itemize}
      \item  total synaptic input current
      \begin{equation}
        \label{eq:all_curr}
        \begin{aligned}
          \text{excitatory neurons:}\quad I_i(t) &= I_{\text{ED},i}(t) + I_{\text{EX},i}(t) + I_{\text{EI},i}(t) ,\ \forall i\in\Epop \\
          \text{inhibitory neurons:}\quad I_i(t) &= I_{\text{IE},i}(t) ,\ \forall i\in\Ipop
        \end{aligned}
      \end{equation}
      with dendritic, inhibitory, excitatory, and external input currents $I_{\text{ED},i}(t)$,  $I_{\text{EI},i}(t)$, $I_{\text{IE},i}(t)$, $I_{\text{EX},i}(t)$ evolving according to
      \begin{equation}
        \label{eq:dendritic_current}
        I_{\text{ED},i}(t)=\sum_{j\in\Epop}(\alpha_{ij}*\xi_j)(t-d_{ij})
        + \sum_{\{k\in\mathcal{T}|i\in\pi_{k,1}\}}(\alpha_{ik}*\xi_k)(t-d_{ik}),
    \end{equation}
    \begin{equation}
      \label{eq:EI_current}
      \tau_\text{EI}\dot{I}_{\text{EI},i} = -I_{\text{EI},i}(t) + \sum_{j\in\Ipop} J_{ij} \xi_j(t-d_{ij}),
    \end{equation}
    \begin{equation}
      \label{eq:IE_current}
      \tau_\text{IE}\dot{I}_{\text{IE},i} = -I_{\text{IE},i}(t) + \sum_{j\in\Epop} J_{ij} \xi_j(t-d_{ij}),
    \end{equation}
    \begin{equation}
      \label{eq:EX_current}
      I_{\text{EX},i}(t)=I_{\text{S},i}(t)+I_{\text{B},i}(t)
    \end{equation}
    with alpha-function kernel $\alpha_{il}(t)=J_{il} \dfrac{e}{\tau_{\text{ED}}} t e^{-t/\tau_{\text{ED}}} \Theta(t)$,\newline
    Heaviside function \mbox{$\Theta(t)=\begin{cases}1 & t \ge 0 \\ 0 & \text{else} \end{cases}$},\newline
    and
    $I_{\text{S},i}(t)$ and $I_{\text{B},i}(t)$ being the stimulus and the background input (see \cref{tab:Model-description-inout}:Input).
    \item suprathreshold dynamics of dendritic currents (dAP generation):
      \begin{itemize}
      \item emission of $k$th dAP of neuron $i$ at time $t_{\text{dAP},i}^k$ if $ I_{\text{ED},i}(t_{\text{dAP},i}^k)\geq\theta_{\text{dAP}}$
      \item dAP current plateau:
      \begin{equation}
        \label{eq:dAP_current_nonlinearity}
        I_{\text{ED},i}(t) = I_\text{dAP}
        \quad\forall{}k,\ \forall t \in \left(t_{\text{dAP},i}^k,t_{\text{dAP},i}^k+\tau_\text{dAP}\right)
      \end{equation}
      with
      dAP current plateau amplitude $I_\text{dAP}$,
      dAP current duration $\tau_\text{dAP}$, and
      dAP activation threshold $\theta_{\text{dAP}}$
      \item reset: $I_{\text{ED},i}(t_{\text{dAP},i}^k+\tau_\text{dAP})=0$ ($\forall{}k$)
      \item reset and refractoriness in response to emission of $l$th somatic spike of neuron $i$ at time $t_{i}^{l}$:
      \begin{equation}
        I_{\text{ED},i}(t)=0
        \quad \forall{}l,\ \forall t \in \left(t_{i}^{l},\,t_{i}^{l}+\tau_{\text{ref},i}\right)
      \end{equation}
      \item[]
    \end{itemize}
    \end{itemize} \\
   \hline 
\end{tabular}
\caption{Description of the synapse model. Parameter values are given in \cref{tab:Model-parameters}.}
\label{tab:Model-description-synapse}
\end{table}
\begin{table}[ht!]
  \small
  \begin{tabular}{|@{\hspace*{1mm}}p{3cm}@{}|@{\hspace*{1mm}}p{12.cm}|}
  \hline 
  \multicolumn{2}{|>{\color{white}\columncolor{black}}c|}{\textbf{Plasticity}}\\
  \hline
  \textbf{Type} & spike-timing dependent structural plasticity and permanence decay \\
  \hline
  \textbf{EE synapses} &
      \begin{itemize}  
        \item dynamics of synaptic permanence $P_{ij}(t)$ (maturity) in EE connections during learning:
          \begin{equation}
            \label{eq:plasticity}
          \begin{aligned}
            \forall P_\text{min}<P_{ij}<P_\text{max}&: \\[1ex]
            \frac{dP_{ij}}{dt} &= P_\text{max} \, \lambda_{+} \sum_{\{t_i^*\}^\prime} \delta(t-[t_i^*+d_\EE]) I_{+}(x_j(t), t_i^*,\Delta{}t_\text{min},\Delta{}t_\text{max}) \\
            &\quad - P_\text{max} \, \lambda_{-} \sum_{\{t_j^*\}^\prime} \delta(t-[t_j^*+d_\EE]) I_{-}(x_i(t),t_j^*,\Delta{}t_\text{max}) \\
            &\quad + (P_\text{min}-P) \frac{1}{\tau_\text{P}}\\
            & \forall{}\{t|P_{ij}(t)<P_\text{min}\}: \quad P_{ij}(t)=P_\text{min}\\
            &\forall{}\{t|P_{ij}(t)>P_\text{max}\}: \quad P_{ij}(t)=P_\text{max}
          \end{aligned}  
        \end{equation}
        with
        \begin{itemize}
        \item list of presynaptic spike times $\{t_j^*\}$,
        \item list of postsynaptic spike times 
          \mbox{$\{t_i^*\}^\prime=\{t_i^*| \forall{}t_j^*:\,t_i^*-t_j^*+d_\EE\ge\Delta{}t_\text{min}\}$}  
        \item increment functions
          \begin{equation}
            \label{eq:indicator_function}
            \begin{aligned}
              I_{+}(x_j(t),t_i^*,\Delta{}t_\text{min},\Delta{}t_\text{max})&=R_{+}(t_i^*-t_j^{+}+d_\EE)\\
              \text{with}\quad
              R_{+}(\tau)&=\
              \begin{cases}
                x_j(t) & \Delta{}t_\text{min}<\tau<\Delta{}t_\text{max}\\
                (x_j(t)-1) & \tau<\Delta{}t_\text{min}\\
                0 & \text{else},
              \end{cases}\\
              I_{-}(x_i(t),t_j^*,\Delta{}t_\text{max})&=R_{-}(t_j^*-t_i^{-}+d_\EE)\\
              \text{with}\quad
              R_{-}(\tau)&=\
              \begin{cases}
                x_i(t) & \tau<\Delta{}t_\text{max}\\
                0 & \text{else},
              \end{cases}
            \end{aligned}
          \end{equation}
        \item maximum permanence $P_\text{max}$, minimum permanence $P_\text{max}$, potentiation and depression rates $\lambda_\text{+}$, $\lambda_\text{-}$, decay time constant $\tau_\text{P}$, delay $d_\EE$, minimum $\Delta{}t_\text{min}$ and maximum $\Delta{}t_\text{max}$ time lags between pairs of pre- and postsynaptic spikes at which synapses are potentiated or depressed, nearest presynaptic spike time $t_j^{+}$ preceding $t_i^*$, nearest postsynaptic spike time $t_i^{-}$ preceding $t_j^*$,
        \item spike trace of presynaptic neuron $j$, evolving according to
        \begin{equation*}
          \frac{dx_j}{dt}=-\tau_{+}^{-1} x_j(t) + \sum_{t_j^*}\delta(t-t_j^*)
        \end{equation*}
        with presynaptic spike times $t_j^*$ and potentiation time constant $\tau_{+}$,
        \item spike trace of postynaptic neuron $i$, evolving according to
        \begin{equation*}
          \frac{dx_i}{dt}=-\tau_{-}^{-1} x_i(t) + \sum_{t_i^*}\delta(t-t_i^*)
        \end{equation*}
        with postynaptic spike times $t_i^*$ and depression time constant $\tau_{-}$.
        \end{itemize}
        \item dynamics of synaptic weights $J_{\EE, ij}$ according to
        \begin{equation*}
            J_{\EE, ij} = \begin{cases}J_\mathrm{max} & P_{ij} > P_\theta \\ 0 & \text{else}, \end{cases}
        \end{equation*}
        with maximum synaptic weight $J_\text{max}$ and synapse maturity threshold $P_\theta$.
      \end{itemize}\\ 
  \hline 
  \textbf{all other synapses} & non-plastic
  \\
  \hline
\end{tabular}
\caption{Description of the plasticity model. Parameter values are given in \cref{tab:Model-parameters}.}
\label{tab:Model-description-plasticity}
\end{table}
\begin{table}
\begin{tabular}{|@{\hspace*{1mm}}p{15.15cm}|}
  \multicolumn{1}{|>{\color{white}\columncolor{black}}c|}{\textbf{Input}}\\
    \begin{itemize}
    \item prediction mode:
    \begin{itemize}
      \item repetitive stimulation of the network using the same
        set $\mathcal{S}=\{s_1,\ldots,s_{S}\}$ of
        sequences $s_i=$\seq{$\zeta_{i,1}$, $\zeta_{i,2}$,\ldots, $\zeta_{i,C_i}$} of
        ordered discrete items $\zeta_{i,j}$ 
        with number of sequences $S$, length $C_i$ of $i$th sequence, and total number of presentations $n_\text{epoch}$
        \item presentation of sequence element $\zeta_{i,j}$ at time $t_{i,j}$ modeled by a single spike $\xi_k(t)=\delta(t-t_{i,j})$ generated by the corresponding external, bottom-up source $\mathcal{X}_k$
        \item generated current in response to the bottom-up presentation of the sequence elements:
        \begin{equation}
          \tau_\text{S}\dot{I}_{\text{S},i} = -I_{\text{S},i}(t) + \sum_{j \in \mathcal{X}} J_{i,j} \xi_j(t-d_{ij})        
        \end{equation} 
    
        \item encoding interval $\Delta{}T=t_{i,j+1}-t_{i,j}$ between subsequent sequence elements $\zeta_{i,j}$ and $\zeta_{i,j+1}$ within a sequence $s_i$
        \item inter-sequence time interval $\Delta{}T_\text{seq}=t_{i+1,1}-t_{i,C_i}$ between subsequent sequences $s_i$ and $s_{i+1}$
        \item top-down inputs $\mathcal{T}_i$ at times $t_{i,1}-\Delta T_\text{td}$ corresponding to the first elements $\zeta_{i,1}$ of all sequences $s_i$ ($\forall{}i \in [1, S]$) to the dendrites of neurons in the respective assembly $\pi_{i,1}$ 
        \item example sequence sets: 
          \begin{itemize}
          \item sequence set I: $\mathcal{S}$=\{\seq{A,F,B,D}, \seq{A,F,C,E}\}
          \item sequence set II: $\mathcal{S}$=\{\seq{A,F,B,D}, \seq{A,F,C,E}, \seq{A,F,G,H}, \seq{A,F,I,J}, \seq{A,F,K,L}\}
          \end{itemize} 
    \end{itemize}
    \item replay mode:
    \begin{itemize}
       \item initiation of replay by stimulation of the minicolumn, corresponding to the first token of the sequence, by a single spike $\xi_k(t)=\delta(t-t_{i,1})$ at time $t_{i,1}$, generated by the corresponding external bottom-up source $\mathcal{X}_k$
       \item top-down inputs $\mathcal{T}_i$ at times $t_{i,1}-\Delta T_\text{td}$ corresponding to the first elements $\zeta_{i,1}$ of all sequences $s_i$ ($\forall{}i \in [1, S]$) to the dendrites of neurons in the respective assembly $\pi_{i,1}$ 
       \item background current $I_\text{B}(t) = I_\text{ext}(t) + I_\text{noise}(t)$ with external input $I_\text{ext}(t)$ and background noise $I_\text{noise}(t)$
       \begin{itemize}
           \item external input:
           \begin{equation}
           I_\text{ext}(t) =
           \begin{cases}
               \bar{I} & \text{(constant input)} \\
               \bar{I} + a\,\text{sin}(2\pi f + \phi) & \text{(oscillatory input)}
           \end{cases}
       \end{equation}
       with offset $\bar{I}$, oscillation amplitude $a$, frequency $f$, and phase $\phi$
       \item noise input:
         \begin{equation}
           \label{eq:shot_noise}
          \tau_\text{noise}\dot{I}_{\text{noise},i} = -I_{\text{noise},i}(t) + J_\text{noise} [x_{\text{E},i}(t) - x_{\text{I},i}(t)]   
        \end{equation} 
        with excitatory and inhibitory Poissonian spike sources $x_{\text{E},i}(t)$ and $x_{\text{I},i}(t)$, synaptic weight $J_\text{noise}$, and time constant $\tau_\text{noise}$
        \vspace{1ex}
        \item[] statistics of background noise:
        \begin{itemize}
            \item noise amplitude
              \begin{equation}
                \sigma = \sqrt{\frac{J_\text{noise}^2 \nu R^2 \tau_\text{noise}^2}{\tau_\text{noise}-\tau_\text{m}}} 
              \end{equation}
              with spike rate $\nu$ of $x_{\text{E},i}(t)$ and $x_{\text{I},i}(t)$, representing standard deviation of the membrane potential of target neurons in the absence other inputs and a spike threshold
            \item noise correlation
              \begin{equation}
                c_{ij} = 
                \begin{cases}
                  c & \text{if } \exists k: i,j \in \mathcal{M}_k\\
                  0 & \text{else}
                \end{cases}
              \end{equation}
              representing correlation coefficient between noise input currents $I_{\text{noise},i}(t)$ and $I_{\text{noise},j}(t)$ 
        \end{itemize}
       \end{itemize}
    \end{itemize}
    \end{itemize}
  \\
\hline 
\end{tabular}
  \caption{Description of the input and the output. Parameter values are given in \cref{tab:Model-parameters}.}
  \label{tab:Model-description-inout}
\end{table}
\begin{table}
\begin{tabular}{|@{\hspace*{1mm}}p{15.15cm}|}
\multicolumn{1}{|>{\color{white}\columncolor{black}}c|}{\textbf{Performance measures}} \\
  \hline
    \begin{itemize}
    \item prediction error:
      \begin{equation}
        \epsilon = \frac{1}{C}\sqrt{\sum_{k=1}^{M}\left(o_k-v_k\right)^2},
      \end{equation}
      with binary vector \mbox{$\boldsymbol{o}\in\{0,1\}^M$}, where $o_k=1$ if subpopulation $k$ is predictive (a minimum of half of assembly size $\rho$ neurons elicit a plateau potential), and $o_k=0$ otherwise. Binary target vector $\boldsymbol{v}$ encodes the identity of the subpopulation receiving the current input
    \begin{itemize}
        \item false positive rate:
          \begin{equation}
            \text{fp} = \frac{1}{C}\sum_{k=1}^{M}\Theta\left(o_k-v_k\right),
          \end{equation}
          quantifies predictive activity in subpopulations not targeted by the current input
        \item false negative rate:          \begin{equation}
            \text{fn} = \frac{1}{C}\sum_{k=1}^{M}\Theta\left(v_k-o_k\right),
          \end{equation}
          quantifies failure to predict the target subpopulation.
    \end{itemize}
        \item replay speed:
    \begin{itemize}
        \item inter-assembly interval $\tau_k$: time between the $k$th and $(k+1)$th assembly activation during replay ($k=1,\ldots,C-1$)
        \item mean inter-assembly interval:
          \begin{equation}
            \langle\tau_k\rangle_k = \frac{1}{4}\sum_{k=C-4}^{C-1}\tau_k,
          \end{equation}
          averaged over the last four elements to exclude transient dynamics at replay onset
        \item replay speed:
          \begin{equation}
            f_\text{out} = \frac{1}{\langle\tau_k\rangle_k},
          \end{equation}
          defined as the inverse of the mean inter-assembly interval, quantifying the rate at which neuronal assemblies are activated during replay
    \end{itemize}
    \item variability of replay timing:
    \begin{itemize}
        \item coefficient of variation of inter-assembly intervals across input oscillation phases:
          \begin{equation}
            \text{CV}_\phi(\tau_k) = \frac{\text{std}_\phi(\tau_k)}{\langle\tau_k\rangle_\phi},
          \end{equation}
          where $\langle\tau_k\rangle_\phi$ and $\text{std}_\phi(\tau_k)$ denote the mean and standard deviation of the $k$th inter-assembly interval across the set of tested oscillation phases $\phi$, respectively; quantifies the sensitivity of the $k$th inter-assembly interval to the oscillation phase at replay onset
        \item coefficient of variation of inter-assembly intervals across sequence elements:
          \begin{equation}
            \text{CV}_k(\tau_k) = \frac{\text{std}_k(\tau_k)}{\langle\tau_k\rangle_k},
          \end{equation}
          where $\langle\tau_k\rangle_k$ and $\text{std}_k(\tau_k)$ denote the mean and standard deviation of the inter-assembly interval across the last four elements $k=C-4,\ldots,C-1$, respectively; quantifies the temporal regularity of replay, i.e., how uniformly the sequence is traversed across successive assembly activations
    \end{itemize}
    \end{itemize} \\
      \hline
\end{tabular}
  \caption{Description of the performance measurement.}
  \label{tab:Performance-measurement}
\end{table}
\begin{table}
\begin{tabular}{|@{\hspace*{1mm}}p{15.15cm}|}
  \multicolumn{1}{|>{\color{white}\columncolor{black}}c|}{\textbf{Initial conditions and network realizations}} \\
    \begin{itemize}
    \item membrane potentials: $V_i(0)=V_\text{r}$ ($\forall{}i\in\mathcal{E}\cup\mathcal{I}$)
    \item dendritic currents: $I_{\text{ED},i}(0)=0$ ($\forall{}i\in\mathcal{E}$)
    \item bottom-up currents: $I_{\text{EX}}(0)=0$ ($\forall{}i\in\mathcal{E}$)
    \item inhibitory currents: $I_{\text{EI},i}(0)=0$ ($\forall{}i\in\mathcal{E}$)
    \item excitatory currents: $I_{\text{IE},i}(0)=0$ ($\forall{}i\in\mathcal{I}$)
    \item synaptic permanences: $P_{ij}(0)\sim\mathcal{U}(P_{0,\text{min}},P_{0,\text{max}})$ (uniform distribution; $\forall{}i,j\in\mathcal{E}$)
    \item synaptic weights: $\J_{ij}(0)=0$ ($\forall{}i,j\in\mathcal{E}$)
    \item spike traces: $x_i(0)=0$ ($\forall{}i\in\mathcal{E}$)
    \item connectivity and initial weights are randomly and independently drawn for each network realization
    \end{itemize}\\
    \hline
  \end{tabular}
\caption{Description of the initial conditions of the network. Parameter values are given in \cref{tab:Model-parameters}.}
\label{tab:Model-description-initcond}
\end{table}
\clearpage
\subsection{Model and simulation parameters}
\label{app:suppl_parameters}
\begin{table}[ht!] 
  \renewcommand{\arraystretch}{1.2}
\begin{tabular}{|@{\hspace*{1mm}}p{3cm}@{}|@{\hspace*{1mm}}p{4cm}@{}|@{\hspace*{1mm}}p{8.1cm}|}
\hline
\textbf{Name} & \textbf{Value} & \textbf{Description}\\
\hline                               
\multicolumn{3}{|>{\columncolor{lightgray}}c|}{\textbf{Network}}\\
\hline 
$M$ & $6$ & number of subpopulations \\
\hline
$\nE$,$\nI$ & $200, 1$ & number of excitatory and inhibitory neurons per subpopulation \\
\hline
\derived{$N_\exc$} & \derived{$1200$} & \derived{number of excitatory neurons; $N_\exc=M\,\nE$} \\
\hline
\derived{$N_\inh$} & \derived{$6$} & \derived{number of inhibitory neurons; $N_\inh=M\,\nI$} \\
\hline
$\rho$ & $20$ & (target) number of active neurons per subpopulation after learning = minimal number of coincident excitatory inputs required to trigger a spike in postsynaptic inhibitory neurons \\
\hline 
$N_\text{X}$ & $6$ & number of bottom-up spike sources; determined by vocabulary size \\
\hline
$N_\text{T}$ & $1$ & number of top-down spike sources; determined by number of training sequences \\
\hline
\multicolumn{3}{|>{\columncolor{lightgray}}c|}{\textbf{(Potential) Connectivity}}\\
\hline
$\KEE$ & $240$ & number of excitatory inputs per excitatory neuron ($\EE$ in-degree) \\
\hline 
$p$ & $K_{\exc\exc}/N_\exc=0.2$ & connection probability \\
\hline 
\hline 
\multicolumn{3}{|>{\columncolor{lightgray}}c|}{\textbf{Excitatory neurons}}\\
\hline 
$\tau_\text{m,E}$ & $10\ms$ & membrane time constant \\
\hline 
$\tau_\text{ref,E}$ & $10\ms$ & absolute refractory period \\
\hline 
$\CM$ & $250\pF$ & membrane capacity \\
\hline 
$\Vreset$ & $0.0\mV$ & reset potential \\
\hline 
$\theta_\text{E}$ & $10\mV$ & somatic spike threshold \\
\hline 
$I_\text{dAP}$ & $200\pA$ &  dAP current plateau amplitude\\
\hline 
$\tau_\text{dAP}$ & $100\ms$ & dAP duration\\
\hline 
$\theta_{\text{dAP}}$ & $59\pA$ & dAP threshold \\
\hline 
\multicolumn{3}{|>{\columncolor{lightgray}}c|}{\textbf{Inhibitory neurons}}\\
\hline
$\tau_\text{m,I}$ & $5\ms$ & membrane time constant\\
\hline 
$\tau_\text{ref,I}$ & $2\ms$ & absolute refractory period\\
\hline 
$\CM$ & $250\pF$ & membrane capacity\\
\hline 
$\Vreset$ & $0.0\mV$ & reset potential\\
\hline 
$\theta_\text{I}$ & $15\mV$ & spike threshold\\
\hline
\end{tabular}
\caption{Model and simulation parameters (continued on next page).}
\label{tab:Model-parameters} 
\end{table}
\setcounter{table}{\thetable-1}
\begin{table}[ht!]
\begin{tabular}{|@{\hspace*{1mm}}p{3cm}@{}|@{\hspace*{1mm}}p{4cm}@{}|@{\hspace*{1mm}}p{8.1cm}|}
\hline
\textbf{Name} & \textbf{Value } & \textbf{Description}\\
\hline
\multicolumn{3}{|>{\columncolor{lightgray}}c|}{\textbf{Synapse}}\\
\hline
$\JIE$ & $\sim{}678.06\pA$ & weight of IE connections (EPSC amplitude) \\
\hline
$\JEI$ & $\sim{}-12915.49\pA$ & weight of EI connections (IPSC amplitude) \\
\hline 
$\JEX$ & $\sim{}4112.20\pA$ & weight of EX connections targeting the somas (EPSC amplitude) \\
\hline
\derived{$\JET$} & \derived{$\sim{}1.4\,\theta_\text{dAP}$} & \derived{weight of ET connections targeting dendrites (EPSC amplitude)} \\
\hline
$J_\text{EN}$ & $\sim{}[0,53.04]\pA$ & weight of EN connections (EPSC amplitude) \\
\hline
${\tau}_{\EE}$ & $5\ms$ & synaptic time constant of EE connections\\
\hline 
${\tau}_{\EI}$ & $1\ms$ & synaptic time constant of EI connections\\
\hline 
${\tau}_{\IE}$ & $0.5\ms$ & synaptic time constant of IE connections\\
\hline
${\tau}_{\EX}$ & $2\ms$ & synaptic time constant of EX connection\\
\hline 
${\tau}_{\ET}$ & $5\ms$ & synaptic time constant of ET connections\\
\hline 
${\tau}_\text{EN}$ & $2\ms$ & synaptic time constant of EN connections\\
\hline
$d_{\EE}$ & $2\ms$ & delay of EE connections (dendritic)\\
\hline
$d_\IE$ & $0.1\ms$ & delay of IE connections\\
\hline
$d_\EI$ & $0.1\ms$ & delay of EI connections\\
\hline 
$d_\EX$ & $0.1\ms$ & delay of EX connections\\
\hline 
$d_{\ET}$ & $2\ms$ & delay of ET connections (dendritic)\\
\hline
\multicolumn{3}{|>{\columncolor{lightgray}}c|}{\textbf{Plasticity}}\\
\hline
$\lambda_{+}$ & $0.39$  & potentiation rate \\
\hline
$\lambda_{-}$ & $0.55$ & depression rate \\
\hline
$P_{ij}$ & $[P_\text{min},P_0]$  & synaptic permanence \\
\hline
$P_{\theta}$ & $20$ & synapse maturity threshold
                            \\
\hline
$P_\text{max}$ & $20$ & permanence upper bound\\
\hline
$P_\text{min}$ & $1$ & permanence lower bound  \\
\hline
$P_{0,\mathrm{max}}$ & $8$ & permanence initialization upper bound \\
\hline
$P_{0,\mathrm{min}}$ & $0$ & permanence initialization upper bound \\
\hline         
$J_\text{max}$ & $12.98\pA$ & maximum weight\\
\hline
$ \tau_{+} $ & $20\ms$ & plasticity time constant (potentiation) \\
\hline
$ \tau_{-} $ & $20\ms$ & plasticity time constant (depression) \\
\hline                        
$ \tau_{\mathrm{P}} $ & $27.5\s$ & permanence leak time constant \\
\hline
$\Delta{}t_\text{min}$ & $4\ms$ & minimum time lag between pairs of pre- and postsynaptic spikes at which synapses are potentiated given the spike trace of the presynaptic neuron of the current time step \\ 
\hline
$\Delta{}t_\text{max}$ & $50\ms$ & maximum time lag between pairs of pre- and postsynaptic spikes at which synapses are potentiated \\
\hline                                             
\multicolumn{3}{|>{\columncolor{lightgray}}c|}{\textbf{Input}}\\
\hline 
$S$ & 1 & number of sequences per sequence set \\
\hline
$C_i$ & $\{8,16,24\}$ & sequence length \\
\hline
$n_\text{epoch}$ & $500$ & total number of sequence set presentations during learning \\
\hline 
$\Delta{}T$ & $40\ms$ & encoding interval (during training)\\
\hline
$\Delta{}T_\text{seq}$ & $\mathcal{U}(100, 105)\ms$ $\qquad \quad$ (uniform distribution) & inter-sequence interval \\
\hline
$\Delta T_\text{td}$ & $10\ms$ & time interval between top-down and bottom-up input \\
\hline
$\nu$ & $10000\,\text{s}^{-1}$ & rate of Poissonian spike sources generating background noise\\
\hline
$\sigma$ & $[0, 3]\mV$ & noise amplitude\\
\hline
$c$ & $0.9$ & noise correlation \\
\hline
\derived{$J_\text{noise}$} & \derived{$[0, 53.04]\pA$} & \derived{synaptic weight of inputs generating background noise; $J_\text{noise} = \sqrt{\frac{\sigma^2\,(\tau_\text{noise}-\tau_\text{m})}{\nu R^2\tau_\text{noise}^2}}$} \\
\hline
\multicolumn{3}{|>{\columncolor{lightgray}}c|}{\textbf{Simulation}}\\
\hline
$\dtsim$ & $0.1\ms$ & time resolution \\
\hline
\end{tabular}
\caption{Model and simulation parameters.}
\end{table}
\subsection{Readout neurons}
{\sloppy
To decode network activity after learning, we employ linear regression to learn a linear mapping from network states to target labels (bottom panels in \cref{fig:slow-sequences}).
To this end, we present all $C$ elements of the sequence to the network once and record the resulting spiking activity of all $\NE$ excitatory neurons.
The network state at each sequence position $c \in \{1,\ldots,C\}$ is represented by a binary vector $\vec{r}_c \in \{0,1\}^{N_\text{E}}$, where the component $i\in[1,N_\text{E}]$ is set to one if neuron $i$ fires a somatic spike within a temporal window of $[-4\,\text{ms}, +14\,\text{ms}]$ relative to the presentation time of the $c$th sequence element, and zero otherwise.
Concatenating these vectors yields the network state matrix
\begin{equation}
    \mat{R} = [\vec{r}_1, \ldots, \vec{r}_C] \in \{0,1\}^{N_\text{E} \times C}.
\end{equation}
The target label for sequence position $c$ is a binary vector $\vec{y}_c \in \{0,1\}^C$ with a one in the $c$th component and zeros elsewhere, encoding the identity of the presented sequence element.
Concatenating these vectors yields the target label matrix
\begin{equation}
    \mat{Y} = [\vec{y}_1, \ldots, \vec{y}_C] = \mat{I}_C \in \{0,1\}^{C \times C},
\end{equation}
which is identical to the $C \times C$ identity matrix $\mat{I}_C$.
Linear regression finds a weight matrix $\mat{W} \in \mathbb{R}^{C \times N_\text{E}}$ such that
\begin{equation}
    \hat{\mat{Y}} = \mat{W}\mat{R} \approx \mat{Y},
\end{equation}
where $\hat{\mat{Y}} \in \mathbb{R}^{C \times C}$ is the decoder matrix.
The components of $\hat{\mat{Y}}$ range from 0 to 1 and quantify the decoding accuracy: the component $\hat{\mat{Y}}_{cc'}$ reflects how confidently the decoder identifies position $c'$ as belonging to sequence element $c$.
Perfect decoding corresponds to $\hat{\mat{Y}} = \mat{I}_C$.
Here, we obtain the weight matrix $\mat{W}$ using \texttt{sklearn.linear\_model.LinearRegression()}~\cite{scikit-learn} via the Functional Neural Architectures (FNA) toolbox~\cite{Duarte21}.
}
\subsection{Hyperparameter tuning}
To obtain optimal learning performance, we perform hyperparameter optimization \cite[Chapter~11]{Goodfellow16} of the synaptic plasticity rule for the most challenging input sequence used throughout the study: the reference melody at one third of the baseline speed, corresponding to a sequence length of $C=24$. We optimize the potentiation rate $\lambda$, depression rate $\lambda_\text{minus}$ and permanence leak time constant $\tau_\text{P}$. The search ranges are $\lambda \in [0.1,0.5]$, $\lambda_\text{minus} \in [0.15, 0.65]$ and $\tau_\text{P} \in [25\s, 50\s]$. 
The hyperparameter tuning uses the Bayesian optimization algorithm of the \textit{Weights \& Biases} framework \cite{Shahriari16,wandb} evaluating learning performance according to the sequence prediction error $\epsilon$ defined in \cref{tab:Performance-measurement}. 
Subsequently, we use the resulting optimal parameter set to train the network on the reference melody at all three learning speeds: baseline, half of baseline, and one third of baseline speed. This approach is motivated by our previous findings \cite{Bouhadjar25_202}, which show that plasticity parameters optimized for longer sequences typically generalize well to shorter sequences, allowing hyperparameter transfer. 
While the sTM model exhibits high theoretical storage capacity \cite{Ahmad16_arXiv,Hawkins16_23}, practical learning efficiency depends on the dynamics of synaptic plasticity.
In particular, the number of training iterations required increases with sequence length \cite{Bouhadjar19_ICONS}, which can make learning long sequences computationally demanding.
\subsection{Simulation details}
All network simulations in this study are performed using the neuronal simulator NEST \cite{Gewaltig_07_11204}, version 3.8 \cite{Nest380}. The excitatory neuron model and the plastic synapse model are defined in the domain-specific language NESTML \cite{Plotnikov16_93,Linssen25_1544143}, version 8.0.0 \cite{Linssen24_12191059}. Simulations are carried out with synchronous updates using exact integration of the system dynamics on a discrete-time grid with a fixed step size $\Delta t$  \cite{Rotter99a}. The source code and simulation data necessary to reproduce the findings of this study are openly available \cite{Lober26_20290700}.
\clearpage
\pdfbookmark[1]{References}{references}

\end{document}